\pdfoutput=1
\documentclass[draft]{agujournal2019}
\draftfalse
\usepackage{url}
\usepackage{lineno}
\usepackage{amsmath}
\usepackage{amssymb}
\usepackage{soul}
\usepackage{epstopdf}

\journalname{JGR-Earth Surface}

\begin{document}
\title{Unified model of sediment transport threshold and rate across weak and intense subaqueous bedload, windblown sand, and windblown snow}
\authors{Thomas P\"ahtz\affil{1}, Yonghui Liu\affil{1}, Yuezhang Xia\affil{1}, Peng Hu\affil{1}, Zhiguo He\affil{1}, and Katharina Tholen\affil{2}}

\affiliation{1}{Institute of Port, Coastal and Offshore Engineering, Ocean College, Zhejiang University, 866 Yu Hang Tang Road, 310058 Hangzhou, China}
\affiliation{2}{Institute for Theoretical Physics, Leipzig University, Br\"uderstra{\ss}e 16, 04103 Leipzig, Germany}

\correspondingauthor{Thomas P\"ahtz}{0012136@zju.edu.cn}
\correspondingauthor{Yuezhang Xia}{yzxia@zju.edu.cn}

\begin{keypoints}
 \item Unified model simultaneously captures sediment transport threshold and rate data across conditions in water and air within a factor of 2
 \item Model supports the recent numerical result that the threshold and rate of equilibrium aeolian saltation are insensitive to soil cohesion
 \item Model challenges the classical prediction that the transport threshold for a longitudinally sloped bed depends on the angle of repose
\end{keypoints}

\begin{abstract}
Nonsuspended sediment transport (NST) refers to the sediment transport regime in which the flow turbulence is unable to support the weight of transported grains. It occurs in fluvial environments (i.e., driven by a stream of liquid) and in aeolian environments (i.e., wind-blown) and plays a key role in shaping sedimentary landscapes of planetary bodies. NST is a highly fluctuating physical process because of turbulence, surface inhomogeneities, and variations of grain size and shape and packing geometry. Furthermore, the energy of transported grains varies strongly due to variations of their flow exposure duration since their entrainment from the bed. In spite of such variability, we here propose a deterministic model that represents the entire grain motion, including grains that roll and/or slide along the bed, by a periodic saltation motion with rebound laws that describe an average rebound of a grain after colliding with the bed. The model simultaneously captures laboratory and field measurements and discrete element method (DEM)-based numerical simulations of the threshold and rate of equilibrium NST within a factor of about 2, unifying weak and intense transport conditions in oil, water, and air (oil only for threshold). The model parameters have not been adjusted to these measurements but determined from independent data sets. Recent DEM-based numerical simulations (Comola, Gaume, et al., 2019, https://doi.org/10.1029/2019GL082195) suggest that equilibrium aeolian NST on Earth is insensitive to the strength of cohesive bonds between bed grains. Consistently, the model captures cohesive windblown sand and windblown snow conditions despite not explicitly accounting for cohesion.
\end{abstract}

\section*{Plain Language Summary}
Loose sedimentary grains cover much of the wind-blown (i.e., aeolian) and water-worked (i.e., fluvial) sedimentary surfaces of Earth and other planetary bodies. To predict how such surfaces evolve in response to aeolian and fluvial flows, one needs to understand the rate at which sediment is transported for given environmental parameters such as the flow strength. In particular, one needs to know the threshold flow conditions below which most sediment transport ceases. Here, we propose a simple model that unifies most aeolian and fluvial sediment transport conditions, predicting both the sediment transport threshold and rate in agreement with measurements and numerical simulations. Our results will make future predictions of planetary surface evolution more reliable than they currently are.

\section{Introduction} \label{Introduction}
When a unidirectional turbulent shearing flow of a Newtonian fluid such as air or water applies a sufficiently strong shear stress $\tau$ onto an erodible sediment bed surface, sediment can be transported by the flow \cite{Duranetal11,Garcia08,Koketal12,Pahtzetal20a,Valanceetal15}. There are two extreme sediment transport regimes: transported grains can enter suspension supported by the flow turbulence and remain out of contact with the bed for very long times, or they can remain in regular contact with the bed (i.e., \textit{nonsuspended}). Fully nonsuspended sediment transport (NST) occurs when the Rouse number $Ro\equiv v^-_s/(\kappa u_\ast)$ exceeds a critical value \cite<$Ro_c\approx2.8$,>[]{Naqshbandetal17}, where $v^-_s$ is the terminal settling velocity of grains in quiescent flow, $\kappa=0.4$ the von K\'arm\'an constant, and $u_\ast\equiv\sqrt{\tau/\rho_f}$ the fluid shear velocity, with $\rho_f$ the fluid density. The most important examples for NST in nature are the transport of coarse sand and gravel driven by streams of liquids (\textit{fluvial}), such as their transport in rivers of water on Earth (\textit{subaqueous bedload}) and rivers of methane on Saturn's moon Titan \cite{Poggialietal16}, and the atmospheric wind-driven (\textit{aeolian}) transport of sand-sized minerals (\textit{windblown sand}) and snow and ice (\textit{windblown snow}).

NST plays a key role in the formation of aeolian and fluvial ripples and dunes on Earth and other planetary bodies \cite{Bourkeetal10,Charruetal13}. Hence, predicting the morphodynamics of planetary sedimentary surfaces requires a deep physical understanding of NST, especially if predictions are to be made outside the range of conditions that are accessible to measurements, like in extraterrestrial environments \cite{ClaudinAndreotti06,Duranetal19,Jiaetal17,Pahtzetal13,Telferetal18}. A first step toward physically understanding NST in its full complexity is to study its most important statistical properties for idealized situations. Numerous physical studies have therefore focused on predicting the sediment transport rate $Q$ (i.e., the total streamwise particle momentum per unit bed area) for equilibrium (i.e., steady and homogeneous) conditions and a flat bed (i.e., no bedforms) of nearly monodisperse, cohesionless, spherical sedimentary grains of density $\rho_p$ and median diameter $d$ \cite<e.g.,>[]{AbrahamsGao06,AliDey17,Ancey20a,Bagnold56,Bagnold66,Bagnold73,Berzietal16,BerziFraccarollo13,Chauchat18,DoorschotLehning02,DuranHerrmann06,Einstein50,FraccarolloHassan19,JenkinsValance14,Lammeletal12,MatousekZrostlik20,Owen64,PahtzDuran20,PomeroyGray90,Sorensen91,Sorensen04,ZankeRoland20}.

Recently, \citeA{PahtzDuran20} unified $Q$ across most aeolian and fluvial environmental conditions (nonshallow flows), parametrized by the gravity constant $g$, bed slope angle $\alpha$ (slopes aligned with flow direction, for downward slopes, $\alpha>0$), kinematic fluid viscosity $\nu_f$, and $\rho_f$ and $\tau$. These authors first defined the following dimensionless numbers:
\begin{subequations}
\begin{linenomath*}
\begin{alignat}{2}
 \textit{Density ratio:}\quad&&s&\equiv\rho_p/\rho_f, \\
 \textit{Galileo number:}\quad&&Ga&\equiv d\sqrt{s\tilde gd}/\nu_f, \\
 \textit{Shields number:}\quad&&\Theta&\equiv\tau/(\rho_p\tilde gd),
\end{alignat}
\end{linenomath*}
\end{subequations}
where $\tilde g\equiv(1-1/s)g\cos\alpha$ is the vertical buoyancy-reduced value of $g$. The Shields number $\Theta$ is a measure for the ratio between tangential flow forces and normal resisting forces acting on bed surface grains \cite{Pahtzetal20a}, $Ga$ controls the scaling of the dimensionless settling velocity $v^-_s/\sqrt{s\tilde gd}$ \cite{Camenen07}, while $\sqrt{s}$ separates the settling velocity scale $\sqrt{s\tilde gd}$ from the escape velocity scale $\sqrt{\tilde gd}$ that grains need to exceed in order to escape the potential traps of the bed surface. \citeA{PahtzDuran20} then separated $Q$ into the transport load $M$ (i.e., the total mass of transported grains per unit bed area) and the average streamwise sediment velocity $\overline{v_x}$ via $Q=M\overline{v_x}$ and derived a parametrization for $Q_\ast\equiv Q/(\rho_pd\sqrt{s\tilde gd})$ that incorporates $M_\ast\equiv M/(\rho_pd)$ and $\overline{v_x}_\ast\equiv\overline{v_x}/\sqrt{s\tilde gd}$ \cite<equation~(\ref{M}) is an improved version of equation~(S20) of>[see \ref{MomentumBalances}]{PahtzDuran20}:
\begin{subequations}
\begin{linenomath*}
\begin{align}
 Q_\ast&=M_\ast\overline{v_x}_{\ast t}(1+c_MM_\ast), \label{Q} \\
 M_\ast&=(\Theta-\Theta_t)/(\mu_b-\tan\alpha), \label{M}
\end{align}
\end{linenomath*}
\end{subequations}
where the subscript $t$ refers to threshold conditions, that is, the limit of vanishing dimensionless transport load, $M_\ast\rightarrow0$ (i.e., $\Theta\rightarrow\Theta_t$, where $\Theta_t$ is the transport threshold).

The parameter $\mu_b$ in equation~(\ref{M}) is the bed surface value of the friction coefficient (i.e., the ratio between particle shear stress and vertical particle pressure), which approximates the ratio between the average streamwise momentum loss and vertical momentum gain of transported grains during their contacts with the bed surface \cite{PahtzDuran18b}. Furthermore, equation~(\ref{Q}) consists of two additive contributions to $Q_\ast$: the term $M_\ast\overline{v_x}_{\ast t}$, which corresponds to the scaling of $Q_\ast$ in the hypothetical case that collisions between transported grains do not occur, and the term $c_MM_\ast^2\overline{v_x}_{\ast t}$, which encodes the effect of such collision on $Q_\ast$. Using discrete element method (DEM)-based numerical simulations of NST, \citeA{PahtzDuran20} found that equation~(\ref{Q}) with $c_M=1.7$ is universally obeyed across simulated weak and intense equilibrium NST conditions ($s\in[2.65,2000]$ and $Ga\in[5,100]$) that satisfy $s^{1/2}Ga\gtrsim80$ for $s\lesssim10$ (typical for fluvial environments) or $s^{1/2}Ga\gtrsim200$ for $s\gtrsim10$ (typical for aeolian environments).

The model of \citeA{PahtzDuran20} is incomplete because it does not incorporate expressions for $\Theta_t$, $\mu_b$, and $\overline{v_x}_{\ast t}$. To overcome this problem, these authors fitted $\Theta_t$ to a given experimental or numerical data set, while they used semiempirical closure relations for $\mu_b$ and $\overline{v_x}_{\ast t}$ from their previous studies: $\mu_b\approx0.63$ \cite{PahtzDuran18b} and $\overline{v_x}_{\ast t}\approx2\kappa^{-1}\sqrt{\Theta_t}$ \cite<limited to $s^{1/4}Ga\gtrsim40$,>[]{PahtzDuran18a}. However, in the context of general modeling, fitting is undesirable, while semiempirical relationships are problematic when applied to conditions outside the range of the data to which they have been adjusted.

Here, we complete the model of \citeA{PahtzDuran20}. Instead of relying on fitting of $\Theta_t$ and semiempirical closure relations for $\mu_b$ and $\overline{v_x}_{\ast t}$, we propose a physical transport threshold model that predicts the three unknown quantities $\Theta_t$, $\mu_b$, and $\overline{v_x}_{\ast t}$ for conditions with arbitrary $s$, $Ga$, and $\alpha$, unifying NST in oil, water, and air. This threshold model is then coupled with equations~(\ref{Q}) and (\ref{M}) to predict $Q_\ast$. We show that this coupled model simultaneously captures laboratory and field measurements and DEM-based numerical simulations of $\Theta_t$ and $Q_\ast$ within a factor of about $2$, though agreement with data of $Q_\ast$ requires that a critical value of $s^{1/2}Ga$ is exceeded (consistent with the validity limitation of equation~(\ref{Q})). The only data of $Q_\ast$ that are not captured by the coupled model (i.e., those with too small $s^{1/2}Ga$) correspond to NST driven by viscous liquids such as oil \cite{Charruetal04}.

From the threshold model predictions, we also derive simple scaling laws for $\Theta_t$ valid for different NST regimes. Special attention is paid to the predicted effect of $\alpha$ on $\Theta_t$ for these regimes, which we compare with the classical bed slope correction of $\Theta_t$.

A further aspect that is addressed in this study is the effect of soil cohesiveness on $\Theta_t$ and $Q_\ast$. From DEM-based numerical simulations of equilibrium aeolian NST for Earth's atmospheric conditions, \citeA{Comolaetal19a} suggested that the strength of cohesive bonds between bed grains does neither significantly affect $\Theta_t$ nor $Q_\ast$ even though it strongly affects the transient toward the equilibrium. However, this suggestion is very controversial. If it was indeed generally true, it would indicate a conceptual problem in most, if not all, existing aeolian transport threshold models. In fact, existing threshold models, regardless of whether they model transport initiation \cite<e.g.,>[]{Burretal15,Burretal20,IversenWhite82,Luetal05,ShaoLu00} or transport cessation \cite<e.g.,>[]{Andreottietal21,Berzietal16,ClaudinAndreotti06,Kok10b,Pahtzetal12}, usually incorporate expressions that describe the entrainment of bed surface grains by the flow and/or grain-bed impacts, and both entrainment mechanisms are strongly hindered by cohesion. Furthermore, even those few existing models that do not consider bed sediment entrainment explicitly account for cohesive forces \cite{Berzietal17,PahtzDuran18a}, which increase the calculated transport threshold.

Here, using numerical data provided by \citeA{Comolaetal19a}, we show that, in contrast to previous threshold models, the conceptualization behind our threshold model supports the suggested insensitivity of aeolian NST to cohesion and propose a criterion for when cohesion can be expected to become important. Consistently, we validate the cohesionless coupled model with transport threshold and rate data not only for cohesionless aeolian and fluvial conditions but also for cohesive aeolian conditions, including aeolian NST of small mineral grains and of potentially very cohesive snow grains.

The reminder of the paper is organized as follows. Section~\ref{Model} presents the transport threshold model, section~\ref{Results} the results, such as the evaluation of the coupled model with existing experimental and numerical data of transport threshold and rate, section~\ref{Discussion} the discussion of the results, and section~\ref{Conclusions} conclusions drawn from it. Furthermore, there are several appendices, containing lengthy justifications of model assumptions and mathematical derivations, and a notations section at the end of the paper.

\section{Transport Threshold Model} \label{Model}
NST is a highly fluctuating physical process \cite{Ancey20b,Duranetal11} because of turbulence, surface inhomogeneities, and variations of grain size and shape and packing geometry. Furthermore, the energy of transported grains varies strongly due to variations of their flow exposure duration since their entrainment from the bed. In fact, grains entrained by the flow initially roll and exhibit a comparably very low kinetic energy, while they can saltate in large hops (especially for aeolian NST), associated with a comparably very large kinetic energy, once they have survived a sufficient number of grain-bed interactions. In spite of such variability, we make the following idealizations to model the transport threshold $\Theta_t$, the bed friction coefficient $\mu_b$, and the threshold value of the dimensionless average streamwise particle velocity $\overline{v_x}_{\ast t}$:
\begin{enumerate}
 \item The mean motion of grains driven by a fluctuating turbulent flow along an inhomogeneous bed surface is represented by the motion of spherical, monodisperse grains driven by the mean (i.e., nonfluctuating) turbulent flow along the mean bed surface (i.e., grain-bed interactions are represented by their statistical mean effect). A partial justification of this idealization is presented in \ref{MeanFlowVelocityAssumption}
 \item Given the idealization above, $\Theta_t$ is defined as the smallest Shields number for which a nontrivial steady state grain trajectory exists. This definition of $\Theta_t$ is conceptually equivalent to the one of \citeA{Almeidaetal06,Almeidaetal08}. It implies that, for $\Theta<\Theta_t$, all transported grains lose more kinetic energy during their rebounds with the bed than they gain during their hops and thus eventually settle
 \item The threshold steady state trajectory is modeled as a periodic saltation trajectory, where grain-bed interactions are described as the average grain-bed rebound, even for NST regimes in which a significant or predominant portion of grains roll and/or roll slide along the bed. However, for this trajectory to be consistent with a sustained rolling motion, we require that grains following this trajectory exhibit a sufficient kinetic energy to be able to roll (or hop) out of the most stable pockets of the bed surface assisted by the near-surface flow. A justification of this idealization is presented in \ref{SaltationJustification}. It is partially based on a bed friction law associated with general equilibrium NST, which is introduced in section~\ref{FrictionLaw}
 \item The quantities $\mu_b$ and $\overline{v_x}_{\ast t}$ are calculated from the periodic saltation trajectory
\end{enumerate}

In the following subsections, we derive step-by-step the transport threshold model. First, we present basic assumptions that characterize flow, particles, and their interactions (section~\ref{BasicSimplifications}). Second, we introduce the bed friction law that equation~(\ref{M}) is based on and show that it leads to an expression linking the average difference between fluid and grain velocity to the bed friction coefficient $\mu_b$ (section~\ref{FrictionLaw}). This friction law, when combined with insights from previous DEM-based numerical simulations of NST, supports representing the entire grain motion in equilibrium NST by grains saltating in identical periodic trajectories with rebound boundary conditions (\ref{SaltationJustification}). Third, we present the mathematical description of this periodic saltation motion (section~\ref{MathematicalModel}). Fourth, we present the manner in which $\Theta_t$, $\mu_b$, $\overline{v_x}_{\ast t}$, and the equilibrium dimensionless sediment transport rate $Q_\ast$ are obtained from the family of identical periodic trajectory solutions (section~\ref{Computation}).

\subsection{Basic Assumptions} \label{BasicSimplifications}
\subsubsection{Flow Velocity Profile} \label{FlowVelocityProfile}
\begin{figure}[htb!]
 \begin{center}
  \includegraphics[width=0.5\columnwidth]{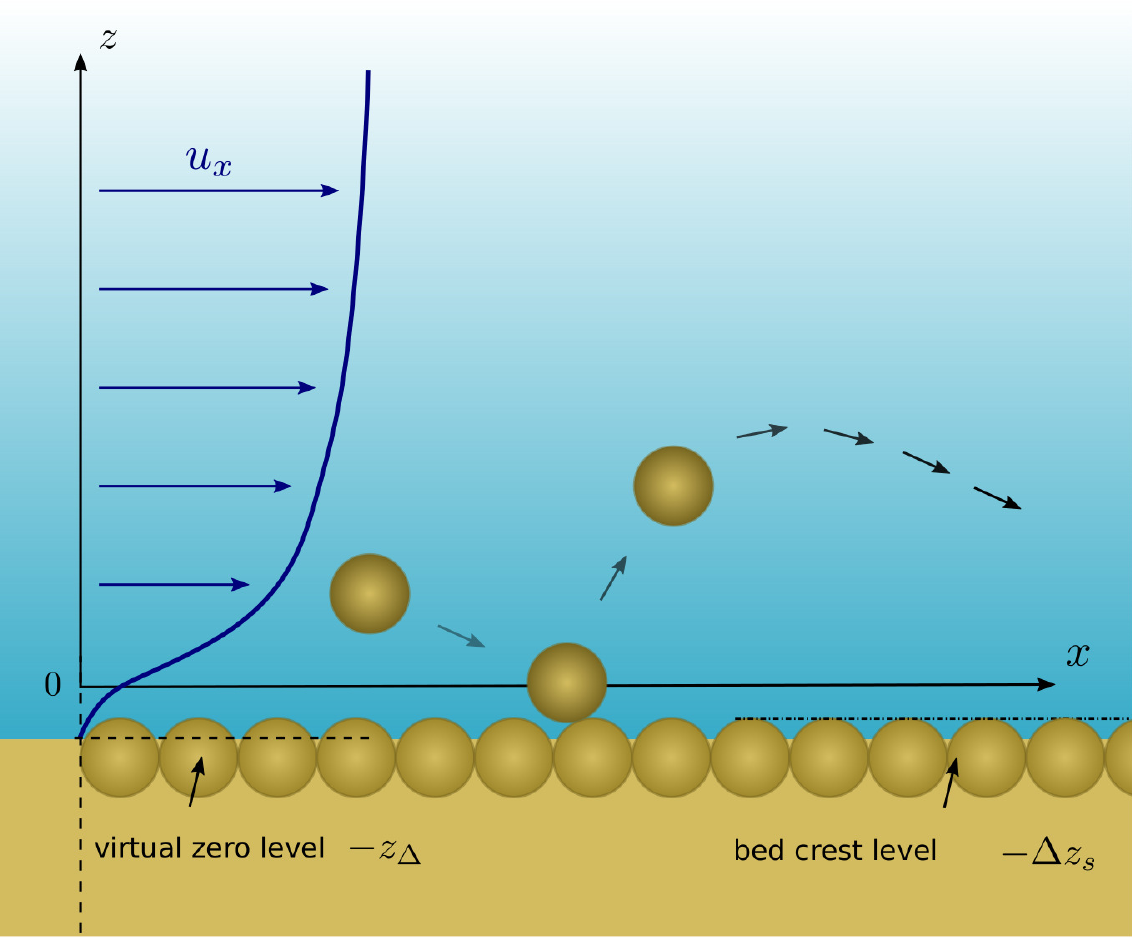}
 \end{center}
 \caption{Sketch visualizing the definition of the bed surface. The bed surface is defined as the elevation at which $p_g\mathrm{d}\langle v_x\rangle/\mathrm{d}z$ is maximal and assigned the vertical coordinate $z=0$, where $p_g$ is the vertical particle pressure, $\mathbf{v}$ the grain velocity, and $\langle\cdot\rangle$ denotes a local grain mass-weighted ensemble average \cite{PahtzDuran18b}. The quantity $\frac{1}{2}p_g\mathrm{d}\langle v_x\rangle/\mathrm{d}z$ describes the production rate of the cross-correlation fluctuation energy density $-\rho\langle v^\prime_zv^\prime_x\rangle$, where $\rho$ is the particle concentration and $\mathbf{v^\prime}\equiv\mathbf{v}-\langle\mathbf{v}\rangle$ the fluctuation grain velocity \cite{PahtzDuran18b}. Since the main source of production of $-\rho\langle v^\prime_zv^\prime_x\rangle$ are grain-bed rebounds, the bed surface elevation defined in this manner corresponds to the effective elevation of the center of mass of energetic grains when they rebound from the bed surface \cite{PahtzDuran18b}. In DEM-based numerical simulations of NST, this elevation is about $z_\Delta=0.7d$ above the virtual zero level of $u_x$ near threshold conditions \cite{PahtzDuran18a}. This value is consistent with measurements of the distance $\Delta z_s\approx0.25d$ between the crest level of static bed grains and the virtual zero level of $u_x$ in both laminar \cite{Hongetal15} and turbulent \cite{Deyetal12} flows. DEM, discrete element method; NST, nonsuspended sediment transport.}
\label{Sketch}
\end{figure}

We consider a mean inner turbulent boundary layer flow above the bed (justified in \ref{MeanFlowVelocityAssumption}) and assume that this flow is undisturbed by the presence of transported grains, since the nondimensionalized mass of transported sediment per unit bed area $M_\ast$ becomes arbitrarily small (i.e., $M_\ast\rightarrow0$) when $\Theta$ approaches $\Theta_t$ from above (i.e., $\Theta\rightarrow\Theta_t$) because of equation~(\ref{M}). The inner turbulent boundary layer is defined by a nearly height-invariant total fluid shear stress in the absence of transported grains \cite<i.e., $\mathrm{d}\tau/\mathrm{d}z\simeq0$,>[]{George13}. This definition implies that the boundary layer thickness or flow depth is much larger than the transport layer thickness (i.e., NST driven by water flows with relatively small flow depth, like in mountain streams, is excluded). The streamwise flow velocity profile $u_x(z)$ within the inner turbulent boundary layer \cite<the \textit{law of the wall},>[]{Smitsetal11} is controlled by the fluid shear velocity $u_\ast$ and the shear Reynolds number $Re_d\equiv u_\ast d/\nu_f=Ga\sqrt{\Theta}$. The law of the wall exhibits three regions: a log layer, $u_x=\kappa^{-1}u_\ast\ln[30(z/d+Z_\Delta)]$, for large wall units $Re_z\equiv Re_d(z/d+Z_\Delta)$; a viscous sublayer, $u_x=u_\ast Re_z$, for small $Re_z$; and a buffer layer for intermediate $Re_z$; where $Z_\Delta\equiv z_\Delta/d=0.7$, with $-z_\Delta$ the virtual zero level of $u_x$ below the bed surface elevation $z=0$ (defined in Figure~(\ref{Sketch})).
The latter two layers vanish when the bed surface becomes too rough (i.e., $Re_d\gtrsim70$). Although it is sometimes conjectured that NST breaks up the viscous sublayer \cite{Koketal12,White79}, DEM-based numerical simulations of NST suggest that this is actually not the case \cite<>[Figure~22]{Duranetal11}. In particular, any potential effect from moving grains should vanish in the limit of threshold conditions because of $M_\ast\rightarrow0$. We use the parametrization of the law of the wall by \citeA{GuoJulien07}, which covers the entire ranges of $Re_z$ and $Re_d$ in a single expression:
\begin{linenomath*}
\begin{equation}
\begin{split}
 u_x&=u_\ast f_u(Re_d,z/d)=u_\ast f_{\tilde u}(Re_d,Re_z),\text{ with} \\
 f_{\tilde u}(Re_d,Re_z)&\equiv7\arctan\left(\frac{Re_z}{7}\right)+\frac{7}{3}\arctan^3\left(\frac{Re_z}{7}\right)-0.52\arctan^4\left(\frac{Re_z}{7}\right) \\
 &+\ln\left[1+\left(\frac{Re_z}{B_\kappa}\right)^{(1/\kappa)}\right]-\frac{1}{\kappa}\ln\left\{1+0.3Re_d\left[1-\exp\left(-\frac{Re_d}{26}\right)\right]\right\}, \label{uxcomplex}
\end{split}
\end{equation}
\end{linenomath*}
where $B_\kappa\equiv\exp(16.873\kappa-\ln9)$.

\subsubsection{Fluid-Particle Interactions} \label{FluidParticleInteractions}
Like recent numerical studies of the physics of aeolian and fluvial sediment transport \cite<e.g.,>[]{Duranetal12,Schmeeckle14}, we consider the fluid drag and buoyancy forces as fluid-particle interactions but neglect other interaction forces because (i) they are usually much smaller than the drag force for grains in motion, (ii) there is no consensus about how these forces behave as a function of the distance from the bed surface, and (iii) we are only looking for the predominant effect and are content with an agreement between model and experimental data within a factor of $2$. Details are explained in \ref{AppendixFluidParticleInteractions}.

\subsubsection{Sedimentary Grains and Sediment Bed} \label{SedimentBed}
We consider a random close packed bed made of nearly monodisperse, cohesionless, spherical sedimentary grains. Furthermore, we assume that flows have worked on this bed for a sufficiently long time such that it has reached a state of maximum resistance \cite{Clarketal17,Pahtzetal20a}. In this state, a quiescent bed surface is able to resist all flows but those whose largest value of the fluctuating fluid shear stress $\tau^{\rm fluc}$, associated with the most energetic turbulent eddy that can possibly form \cite<note that the size, and thus the energy content, of eddies is limited by the system dimensions,>[]{Smitsetal11}, exceeds a certain critical resisting shear stress $\Theta_Y\rho_p\tilde gd$ \cite{Clarketal17}. In particular, for laminar (i.e., nonfluctuating) fluvial conditions ($\tau^{\rm fluc}=\tau$), such a bed surface is able to resist all driving flows with Shields numbers $\Theta<\Theta_Y$ \cite{Pahtzetal20a}. The so-called \textit{yield stress} $\Theta_Y$ is therefore a statistical quantity encoding the resistance of the bed surface as a whole, though it may be interpreted as the nondimensionalized fluctuating fluid shear stress $\Theta^{\rm fluc}\equiv\tau^{\rm fluc}/(\rho_p\tilde gd)$ that is required to initiate rolling of grains resting in the most stable pockets of the bed surface \cite{Clarketal17}. For a nonsloped bed of nearly monodisperse, cohesionless, frictional spheres, $\Theta^o_Y\equiv\Theta_Y|_{\alpha=0}$ is expected to exhibit a universal value \cite{Pahtzetal20a}. Based on measurements for laminar fluvial driving flows \cite{Charruetal04,Houssaisetal15,Loiseleuxetal05,Ouriemietal07}, we use the approximate value $\Theta^o_Y=0.13$.

\subsection{Bed Friction Law} \label{FrictionLaw}
The transport threshold model derivation starts with describing general equilibrium NST by a bed friction law that goes back to \citeA{Bagnold56}; \citeA{Bagnold66,Bagnold73}, and which also led to the derivation of equation~(\ref{M}). In fact, the bed friction coefficient $\mu_b$ in equation~(\ref{M}) is rigorously linked to the streamwise ($a_x$) and vertical ($a_z$) components of the acceleration $\mathbf{a}$ of transported grains due to noncontact forces via \cite{PahtzDuran18a,PahtzDuran20}
\begin{linenomath*}
\begin{equation}
 \mu_b=-\overline{a_x}/\overline{a_z}, \label{mubdef}
\end{equation}
\end{linenomath*}
where the overbar denotes the particle concentration ($\rho$)-weighted height average, $\overline{\cdot}\equiv\frac{1}{M}\int_0^{z_{\rm max}}\rho\langle\cdot\rangle\mathrm{d}z$, with $z_{\rm max}$ the top of the transport layer and $M\equiv\int_0^{z_{\rm max}}\rho\mathrm{d}z$ (the same $M$ as $M=\rho_pdM_\ast$ in equation~(\ref{M})). The grain acceleration $\mathbf{a}$ consists of a drag (superscript $d$), a gravity (superscript $g$), and a buoyancy (superscript $b$) component. In particular, $a_z=a^d_z+a^g_z+a^b_z=a^d_z-\tilde g$. Hence, defining $S\equiv-(a^g_x+a^b_x)/(a^g_z+a^b_z)$, $\mathbf{a}$ can be expressed as
\begin{linenomath*}
\begin{equation}
 (a_x,a_z)=(a^d_x+S\tilde g,a^d_z-\tilde g). \label{axdef}
\end{equation}
\end{linenomath*}
Note that $S\simeq(1-1/s)^{-1}\tan\alpha$ for slope-driven NST in turbulent liquids and $S\simeq\tan\alpha$ for aeolian NST and slope-driven NST in viscous liquids \cite<these differences arise because $\mathbf{a^b}$ is proportional to the divergence of only the viscous contribution to the fluid stress tensor,>[]{Maurinetal18}. Since these conditions cover most natural environments, $S$ is treated as a further constant dimensionless number (the \textit{slope number}) characterizing a given NST threshold condition in addition to $Ga$ and $s$.

In order to allow for an easy analytical evaluation of equation~(\ref{mubdef}), we linearize $\mathbf{a^d}$ via approximating the difference $|\mathbf{u}-\mathbf{v}|$ between fluid ($\mathbf{u}$) and grain ($\mathbf{v}$) velocity by the mean value of its streamwise component: $|\mathbf{u}-\mathbf{v}|\approx\overline{u_x}-\overline{v_x}$. We carried out a few tests with the final transport threshold model that suggested that this approximation has almost no effect on the final prediction. Using a standard drag law for spherical grains \cite{Camenen07,FergusonChurch04}, the linearized drag acceleration reads
\begin{linenomath*}
\begin{equation}
 \mathbf{a^d}=\frac{3}{4sd}\left[\sqrt[m^d]{\frac{24\sqrt{s\tilde gd}}{Ga}}+\sqrt[m^d]{C_d^\infty\left(\overline{u_x}-\overline{v_x}\right)}\right]^{m^d}(\mathbf{u}-\mathbf{v}), \label{ad}
\end{equation}
\end{linenomath*}
where $C^\infty_d=0.4$ and $m^d=2$ \cite<for nonspherical grains, different values of $C^\infty_d$ and $m^d$ are more appropriate,>[]{Camenen07}. Equation~(\ref{ad}) implies two mathematical identities. First, $\overline{a^d_x}/\tilde g=(\overline{u_x}-\overline{v_x})/v_s$, where $v_s\equiv-v_z|_{a^d_z=\tilde g}$ is the terminal grain settling velocity, because of $a^d_x/a^d_z=(u_x-v_x)/(u_z-v_z)$ and $u_z=0$. Second, $\overline{a^d_z}=0$ because of $\overline{v_z}=0$ \cite<mass conservation,>[]{Pahtzetal15a} and $u_z=0$. Using equations~(\ref{mubdef}) and (\ref{axdef}), $\overline{a^d_z}=0$ further implies $\overline{a^d_x}/\tilde g=\mu_b-S$. After using the latter relation in equation~(\ref{ad}) and taking the $m^d$-th root, it follows that $[(\overline{u_x}-\overline{v_x})/\sqrt{s\tilde gd}]^{1/m^d}$ obeys a quadratic equation with coefficients that depend on $\mu_b-S$ and $Ga$. Solving this quadratic equation and combining it with the first mathematical identity, we obtain the following expression for $\overline{u_x}-\overline{v_x}$ and the dimensional (nondimensionalized) settling velocity $v_s$ ($v_{s\ast}$):
\begin{linenomath*}
\begin{equation}
\begin{split}
 v_{s\ast}&\equiv\frac{v_s}{\sqrt{s\tilde gd}}=\frac{\overline{u_x}-\overline{v_x}}{(\mu_b-S)\sqrt{s\tilde gd}} \\
 &=\frac{1}{\mu_b-S}\left[\sqrt{\frac{1}{4}\sqrt[m^d]{\left(\frac{24}{C_d^\infty Ga}\right)^2}+\sqrt[m^d]{\frac{4(\mu_b-S)}{3C_d^\infty}}}-\frac{1}{2}\sqrt[m^d]{\frac{24}{C_d^\infty Ga}}\right]^{m^d}. \label{Settling}
\end{split}
\end{equation}
\end{linenomath*}
A similar link between $\overline{u_x}-\overline{v_x}$ and $v_s$ as in equation~(\ref{Settling}) was previously established by \citeA{Bagnold73}, while the expression for $\overline{u_x}-\overline{v_x}$ has been validated with data from DEM-based numerical simulations of NST for a wide range of conditions \cite{PahtzDuran18a}. Both facts support using a linearized drag law (equation~(\ref{ad})). Note that the terminal grain settling velocity in quiescent flow $v^-_s$, which is distinct from $v_s$, obeys a modified version of the lower line of equation~(\ref{Settling}) in which $\mu_b-S$ is replaced by $1$ \cite{Camenen07}.

Equation~(\ref{Settling}), when combined with insights from previous DEM-based numerical simulations of NST into the physical nature of $\mu_b$, supports representing the entire grain motion in equilibrium NST by grains saltating in identical periodic trajectories with rebound boundary conditions (\ref{SaltationJustification}).

\subsection{Mathematical Description of Periodic Saltation} \label{MathematicalModel}
This subsection introduces the mathematical description of the main model idealization: saltation in identical periodic trajectories along a flat wall. Equation~(\ref{Settling}) is a part of this description. However, it is not immediately clear what are the physical meanings of $\mu_b$ and the concentration ($\rho$)-weighted height average $\overline{\cdot}$, both appearing in equation~(\ref{Settling}), in the context of periodic saltation. Below, we therefore derive expressions for both items in terms of three quantities that are intuitively associated with periodic saltation: the lift-off or rebound velocity $\mathbf{v_\uparrow}$, impact velocity $\mathbf{v_\downarrow}$, and hop time $T$.

In periodic saltation, a grain crosses each elevation $z<H$, where $H$ is the hop height, twice during a single saltation trajectory: once for times before (superscript $\uparrow$) and once for times after (superscript $\downarrow$) the instant $t(H)$ at which it reaches its highest elevation $z=H$. In particular, due to mass conservation, the upward vertical mass flux of grains $\phi^\uparrow$, which is equal to the negative downward vertical mass flux of grains, is constant for $z<H$, while $\phi^\uparrow=0$ for $z>H$ \cite{Berzietal16}. Hence, for $z<H$, the concentrations of ascending and descending grains are given by $\rho^\uparrow(z)=\phi^\uparrow/v^\uparrow_z(z)$ and $\rho^\downarrow(z)=-\phi^\uparrow/v^\downarrow_z(z)$, respectively (note that $\mathbf{v^{\uparrow(\downarrow)}}(0)=\mathbf{v_{\uparrow(\downarrow)}}$). Using $\rho(z)=\rho^\uparrow(z)+\rho^\downarrow(z)$, it follows that the $\rho$-weighted height average of a grain quantity $G$ is equal to its time average:
\begin{linenomath*}
\begin{equation} 
 \overline{G}\equiv\frac{\int_0^{z_{\rm max}}\rho\langle G\rangle\mathrm{d}z}{\int_0^{z_{\rm max}}\rho\mathrm{d}z}=\frac{\phi^\uparrow\int_0^H\left(\frac{G^\uparrow}{v^\uparrow_z}-\frac{G^\downarrow}{v^\downarrow_z}\right)\mathrm{d}z}{\phi^\uparrow\int_0^H\left(\frac{1}{v^\uparrow_z}-\frac{1}{v^\downarrow_z}\right)\mathrm{d}z}=\frac{\int_0^{t(H)}G^\uparrow\mathrm{d}t-\int_T^{t(H)}G^\downarrow\mathrm{d}t}{\int_0^{t(H)}\mathrm{d}t-\int_T^{t(H)}\mathrm{d}t}=\frac{1}{T}\int_0^TG\mathrm{d}t.
\end{equation}
\end{linenomath*}
Furthermore, using $\mu_b=-\langle v^\prime_zv^\prime_x\rangle(0)/\langle v_z^{\prime2}\rangle(0)$ (\ref{SaltationJustification}) and $\langle v^\prime_zv^\prime_i\rangle=\langle v_zv_i\rangle$ because of $\langle v_z\rangle=0$ \cite<mass conservation,>[]{Pahtzetal15a}, it also follows that one can link $\mu_b$ to the rebound and impact velocities:
\begin{linenomath*}
\begin{equation} 
 \mu_b=-\frac{[\rho^\uparrow v^\uparrow_zv^\uparrow_x](0)+[\rho^\downarrow v^\downarrow_zv^\downarrow_x](0)}{[\rho^\uparrow v_z^{\uparrow2}](0)+[\rho^\downarrow v_z^{\downarrow2}](0)}=\frac{v_{\downarrow x}-v_{\uparrow x}}{v_{\uparrow z}-v_{\downarrow z}}. \label{mub}
\end{equation}
\end{linenomath*}

Now, we subdivide this subsection into further subsections. First, we present the deterministic laws governing the motion of a grain above the bed driven by the mean turbulent flow (section~\ref{IdenticalTrajectories}). These laws directly map $\mathbf{v_\uparrow}$ to $\mathbf{v_\downarrow}$. Second, we present the laws describing grain-bed rebounds (section~\ref{ParticleBedRebounds}), mapping $\mathbf{v_\downarrow}$ back to $\mathbf{v_\uparrow}$. For the grain trajectories to be identical and periodic, these laws must also be deterministic, which is achieved by representing them by their statistical mean effect. Third, we model the critical Shields number $\Theta^{\rm roll}$ needed for a grain of a given kinetic energy, associated with a given periodic saltation trajectory, to roll (or hop) out of the most stable pockets of the bed surface assisted by the near-surface flow (section~\ref{ReboundEnergy}). For a periodic saltation trajectory to be consistent with a sustained rolling motion, we require that $\Theta>\Theta^{\rm roll}$.

\subsubsection{Grain Motion Above the Bed} \label{IdenticalTrajectories}
To make the analytical notation compact, we nondimensionalize location, velocity, acceleration, and time, indicated by a hat, using combinations of the terminal settling velocity and reduced gravity: $v_s^2/\tilde g$, $v_s$, $\tilde g$, and $v_s/\tilde g$, respectively. Using $\hat z\equiv z/(v_s^2/\tilde g)=z/(v_{s\ast}^2sd)$, one then obtains the following system of differential equations describing the average trajectory from equations~(\ref{uxcomplex}), (\ref{axdef}), and (\ref{ad}):
\begin{linenomath*}
\begin{subequations}
\begin{align}
 \hat u_x(\hat z)&=v_{s\ast}^{-1}\sqrt{\Theta}f_u\left(Ga\sqrt{\Theta},sv_{s\ast}^2\hat z\right), \label{LawWall} \\
 \frac{\mathrm{d}}{\mathrm{d}\hat t}\hat v_x&=\hat u_x-\hat v_x+S \label{ax}, \\
 \frac{\mathrm{d}}{\mathrm{d}\hat t}\hat v_z&=-1-\hat v_z \label{az}.
\end{align}
\end{subequations}
\end{linenomath*}
The solution of equations~(\ref{LawWall})-(\ref{az}), with the initial condition $\mathbf{\hat v}(0)=\mathbf{\hat v_\uparrow}$, is straightforward and given in \ref{AppendixAnalyticalSolution}. For the transport threshold model, the following expressions, which can be obtained from the solution (\ref{AppendixAnalyticalSolution}), are crucial (written in a form that allows easy iterative evaluation, see section~\ref{Computation}):
\begin{linenomath*}
\begin{align}
 \hat v_{\downarrow z}&=-1-W\left[-(1+\hat v_{\uparrow z})e^{-(1+\hat v_{\uparrow z})}\right], \label{viz} \\
 \Theta&=\frac{\sqrt{\Theta}v_{s\ast}[\mu_b(1+\hat v_{\uparrow z})+\hat v_{\uparrow x}-S]}{f_u\left\{Ga\sqrt{\Theta},sv_{s\ast}^2\left[-\hat v_{\downarrow z}(1+\hat v_{\uparrow z})-\hat v_{\uparrow z}\right]\right\}}, \label{Thetacompapp}
\end{align}
\end{linenomath*}
where $W$ denotes the principal branch of the Lambert-$W$ function.

\subsubsection{Grain-Bed Rebounds} \label{ParticleBedRebounds}
\begin{figure}[htb!]
 \begin{center}
  \includegraphics[width=1.0\columnwidth]{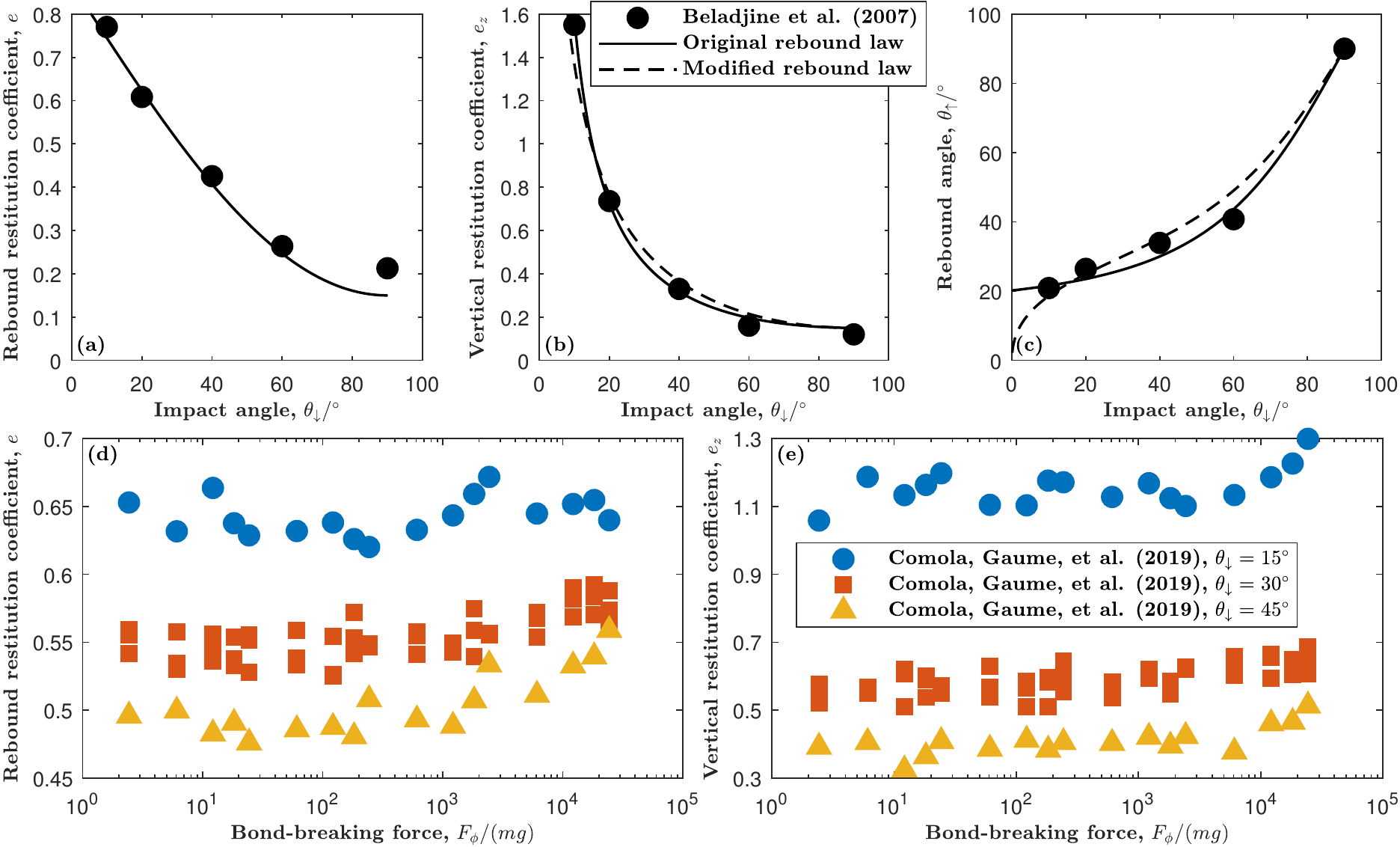}
 \end{center}
 \caption{Rebound laws. (a) Rebound restitution coefficient $e$, (b) vertical restitution coefficient $e_z$, and (c) rebound angle $\theta_\uparrow=\arcsin(e_z\sin\theta_\downarrow/e)$ versus impact angle $\theta_\downarrow$. (d) Rebound restitution coefficient $e$ and (e) vertical restitution coefficient $e_z$ versus the dimensionless force $F_\phi/(mg)$ needed to break cohesive bonds between bed grains. Symbols in (a)-(c) correspond to averaged experimental data for nearly monodisperse, cohesionless, spherical grains ($\rho_p=1770~\mathrm{kg/m}^3$ and  $d=6~\mathrm{mm}$) \cite{Beladjineetal07}. The solid line in (a) corresponds to equation~(\ref{e}). The solid (dashed) lines in (b) and (c) correspond to equation~(\ref{ezoriginal}) (equation~(\ref{ez})). Symbols in (d) and (e) correspond to averaged DEM numerical simulation data by \citeA{Comolaetal19a} for $\rho_p=1000~\mathrm{kg/m}^3$, $g=9.81~\mathrm{m/s}^2$, $m=\rho_p\pi d^3/6$, three different impact velocities $|\mathbf{v_\downarrow}|=(1.58,3.16,4.74)~\mathrm{m/s}$, and three different impact angles $\theta_\downarrow=(15^\circ,30^\circ,45^\circ)$. The rebounding spherical grain has a diameter of $d=200~\mu$m, while the diameter of the spherical bed grains obeys a log-normal distribution with mean $\overline{d}=200~\mu\mathrm{m}$ and standard deviation $\sigma_d=50~\mu\mathrm{m}$. The figure legend in (d) is the same as in (e). Figure legends in (a) and (c) are the same as in (b). DEM, discrete element method.}
\label{Rebound}
\end{figure}

Grain collisions with a static sediment bed have been extensively studied experimentally \cite{Ammietal09,Beladjineetal07}, numerically \cite{Comolaetal19a,Lammeletal17,Tanabeetal17}, and analytically \cite{ComolaLehning17,Lammeletal17}. In typical experiments, an incident grain is shot with a relatively high impact velocity ($|\mathbf{v_\downarrow}|\gg\sqrt{\tilde gd}$) onto the bed and the outcome of this impact (i.e., the grain rebound and potentially ejected bed grains) statistically analyzed. We describe this process using a phenomenological description for the average rebound (vertical) restitution coefficient $e\equiv|\mathbf{v_\uparrow}|/|\mathbf{v_\downarrow}|$ ($e_z\equiv-v_{\uparrow z}/v_{\downarrow z}$) as a function of $\sin\theta_\downarrow=-v_{\downarrow z}/|\mathbf{v_\downarrow}|$, with $\theta_\downarrow$ the average impact angle \cite{Beladjineetal07}:
\begin{linenomath*}
\begin{subequations}
\begin{alignat}{2}
 &&e&=A-B\sin\theta_\downarrow, \label{e} \\
 \text{Original:}\quad&&e_z&=A_z/\sin\theta_\downarrow-B_z, \label{ezoriginal} \\
 \text{Modified:}\quad&&e_z&=(A+C)/\sqrt{\sin\theta_\downarrow}-(B+C), \label{ez}
\end{alignat}
\end{subequations}
\end{linenomath*}
where $A=0.87$, $B=0.72$, $A_z=0.3$, $B_z=0.15$, and $C=0$. Equation~(\ref{ez}) is our modification of equation~(\ref{ezoriginal}), the original expression given by \citeA{Beladjineetal07}. This modification accounts for the analytically derived asymptotic behavior of the rebound angle in the limit of small impact angle, $\sin\theta_\uparrow=v_{\uparrow z}/|\mathbf{v_\uparrow}|=e_z\sin\theta_\downarrow/e\sim\sqrt{\sin\theta_\downarrow}$ \cite{Lammeletal17}, and for the requirement that $\theta_\uparrow\rightarrow90^\circ$ when $\theta_\downarrow\rightarrow90^\circ$. Like the original expressions, the modified expressions are consistent with experimental data by \citeA{Beladjineetal07} for nearly monodisperse, cohesionless, spherical grains, as shown in Figures~\ref{Rebound}(a)-\ref{Rebound}(c).

We assume that equations~(\ref{e}) and (\ref{ez}) are roughly universal for monodisperse, cohesionless, spherical grains, independent of $|\mathbf{v_\downarrow}|$ and bed-related parameters, such as $\rho_p$, $\tilde g$, and $d$, since the experimental data by \citeA{Beladjineetal07} have also been reproduced by a theoretical model that predicts the rebound parameters as a function of only $\theta_\downarrow$ \cite{Lammeletal17}. Furthermore, for conditions in which vertical drag is negligible (i.e., $|a^d_z|\ll\tilde g$, $e_z\simeq1$, $\overline{v_x}\simeq(v_{\uparrow x}+v_{\downarrow x})/2$, and $\overline{v_z^2}\simeq v_{\uparrow z}^2/3$, see \ref{AppendixNoVerticalDrag}), any given set of rebound laws that depends only on $\theta_\downarrow$, such as equations~(\ref{e}) and (\ref{ez}), results in $v_{\uparrow z}=-v_{\downarrow z}\propto v_{\uparrow x}\propto v_{\downarrow x}$ with fixed proportionality constants, implying $\mu_b=\mathrm{const}$ (equation~(\ref{mub})) and $(\overline{v_z^2})^{1/2}\propto\overline{v_x}$. Qualitatively, these two scaling relations correspond to the semiempirical closures used in the threshold model of \citeA{PahtzDuran18a} (equations~(\ref{mubt}) and (\ref{vzvx})), based on which we motivate the usage of universal rebound boundary conditions across all NST regimes in \ref{SaltationJustification}.

Figures~\ref{Rebound}(d) and \ref{Rebound}(e) show that, for the DEM numerical simulation data by \citeA{Comolaetal19a}, $e$ and $e_z$ are insensitive to the cohesiveness of the bed material. In fact, for a bed consisting of spherical grains with log-normally distributed size and a grain impacting with a relatively high impact velocity ($|\mathbf{v_\downarrow}|\gg\sqrt{\tilde gd}$), these authors varied the critical force $F_\phi$ that is required to break cohesive bonds between bed grains over several orders of magnitude and found nearly no effect on the average rebound dynamics. This finding will play a crucial role in section~\ref{Cohesion}, where we discuss the importance of cohesion for NST.

Lastly, we note that viscous damping of binary collisions \cite<e.g.,>[]{Gondretetal02}, which can be important for fluvial NST, also does not seem to significantly affect the rebound laws \cite<>[section 4.1.1.4]{Pahtzetal20a}, in contrast to the assumptions in previous trajectory-based transport threshold models \cite{Berzietal16,Berzietal17}. In particular, DEM-based simulations of aeolian and fluvial NST indicate that, even for nearly fully damped binary collisions (normal restitution coefficient $\epsilon=0.01$), grain-bed rebounds (and thus $e$ and $e_z$) are not much affected when compared with nondamped ($\epsilon=0.9$) binary collisions \cite<>[see particularly their Movies~S1-S3]{PahtzDuran18a}. This probably means that the tangential relative velocity component, which is not much affected by $\epsilon$, dominates during the rebound process.

\subsubsection{Critical Shields Number Required for Rolling} \label{ReboundEnergy}
\begin{figure}[htb!]
 \begin{center}
  \includegraphics[width=0.5\columnwidth]{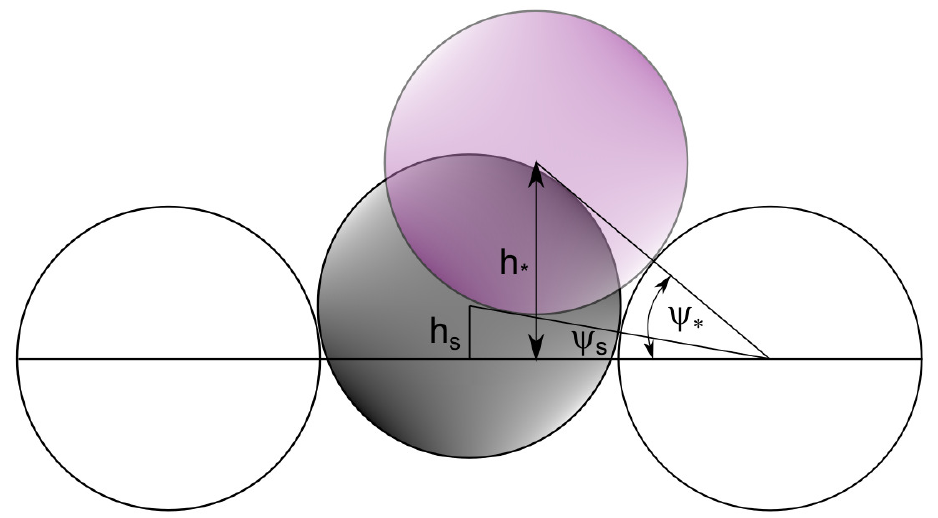}
 \end{center}
 \caption{Sketch of a grain of a given kinetic energy $E_\uparrow$ rolling out of the most stable bed surface pocket. The process is modeled as follows. The grain first rolls along its downstream neighbor (neglecting rolling friction and flow driving) until $E_\uparrow$ has been fully converted into potential energy $m\tilde g(h_\ast-h_s)$, increasing the pocket angle via $\sin\psi_\ast=\sin\psi_s+E_\uparrow/(m\tilde gd)$. A sufficiently strong flow can then push the grain out from this new location within the pocket.}
\label{Escape}
\end{figure}

In \ref{SaltationJustification}, we provide justifications for why one can represent the entire grain motion in NST, including grains that roll and/or slide along the bed, by a pure saltation motion. However, to be consistent with a sustained rolling motion of grains, we require that saltation trajectories are limited to Shields numbers $\Theta$ that are larger than the critical value $\Theta^{\rm roll}$ needed for a grain of a given kinetic energy, associated with a given periodic saltation trajectory, to roll (or hop) out of the most stable pockets of the bed surface assisted by the near-surface flow. In this section, we derive an expression for $\Theta^{\rm roll}$ using a highly simplified approach. First, since a grain located in such a pocket just changed its direction of motion from downward to upward, we assume that it exhibits the rebound kinetic energy $E_\uparrow=\frac{1}{2}m\mathbf{v_\uparrow}^2$, where $m$ is the grain mass, of the given periodic saltation trajectory. Second, we assume that this grain first rolls along its downstream neighbor until $E_\uparrow$ has been fully converted into potential energy $m\tilde g(h_\ast-h_s)$ (Figure~\ref{Escape}), neglecting rolling friction and flow driving. 
This rolling motion increases the pocket angle from the value $\psi_s$ corresponding to the most stable bed surface pocket to the value $\psi_\ast$ via (Figure~\ref{Escape})
\begin{linenomath*}
\begin{equation}
 \sin\psi_\ast=\sin\psi_s+E_\uparrow/(m\tilde gd). \label{phiast}
\end{equation}
\end{linenomath*}
We then model $\Theta^{\rm roll}$ as the critical Shields number required to push the grain out from this new pocket angle position, assuming that the mean grain motion driven by a turbulent flow is the same as the mean grain motion driven by the mean turbulent flow (the first idealization in section~\ref{Model}).

Our definition of $\Theta^{\rm roll}$ implies that $\Theta^{\rm roll}$ depends on $E_\uparrow/(m\tilde gd)$ via equation~(\ref{phiast}) and vanishes for sufficiently large values of $E_\uparrow/(m\tilde gd)$, for which the grain can roll (or hop) out of the pocket all by itself (i.e., for $\psi_\ast\geq90^\circ$). Furthermore, in the limit $E_\uparrow/(m\tilde gd)\rightarrow0$ (i.e., $\psi_\ast\rightarrow\psi_s$), $\Theta^{\rm roll}\rightarrow\Theta_Y$, since the yield stress $\Theta_Y$ can be interpreted as the Shields number required to initiate rolling of grains resting in the most stable pockets of the bed surface in the absence of turbulent fluctuations (section~\ref{SedimentBed}). This limit is relevant for $Ga^2s\lesssim1$, typical for NST driven by laminar fluvial flows \cite{PahtzDuran18a}, where the average energy of transported grains scales with $Ga^2sm\tilde gd$ \cite{Charruetal04} and thus $E_\uparrow/(m\tilde gd)$ is relatively small.

For a nonsloped ($S=0$) bed of triangular or quadratic geometry and a laminar driving flow, \citeA{Agudoetal17} derived nearly exact analytical expressions for the critical Shields number $\Theta_c$ required to push a grain out from an arbitrary pocket angle position $\psi$. As triangular arrangements are the most probable ones in disordered configurations \cite{Agudoetal17}, we assume that these expressions approximately apply also to natural sediment beds. For the special case $\psi=25^\circ$, these expressions predict $\Theta_c|_{\psi=25^\circ}\simeq0.13$. Since the nonsloped yield stress exhibits the same value, $\Theta^o_Y=0.13$ (section~\ref{SedimentBed}), we identify this pocket angle as the one of the most stable bed surface pocket (i.e., $\psi_s=25^\circ$). \citeA{Agudoetal17} further noted that their analytical expressions are reasonably well approximated by the nonsloped version of the model of \citeA{WibergSmith87}: $\Theta_c\propto\cot\psi$ for $S=0$. Assuming that the general version of the model of \citeA{WibergSmith87}, $\Theta_c\propto\cot\psi-S$, works reasonably well for arbitrarily sloped beds \cite<which have not been studied by>[]{Agudoetal17}, we conclude that $\Theta_c=(\cot\psi-S)\Theta^o_Y/\cot\psi_s$. Hence, identifying $\Theta^{\rm roll}$ as $\Theta_c|_{\psi=\psi_\ast}$, using equation~(\ref{phiast}) with $E_\uparrow/(m\tilde gd)=\frac{1}{2}sv_{s\ast}^2\mathbf{\hat v_\uparrow}^2$, and imposing that $\Theta^{\rm roll}=0$ when the grain exhibits a sufficient kinetic energy to be able to roll (or hop) out the most stable bed surface pocket all by itself, we obtain
\begin{linenomath*}
\begin{equation}
\begin{split}
 \Theta^{\rm roll}&=\frac{\Theta^o_Y}{\cot\psi_s}\max\left[\cot\psi_\ast-S\text{sgn}(\cot\psi_\ast),0\right],\text{ with} \\
 \cot\psi_\ast&=\sqrt{\max\left[\left(\sin\psi_s+\frac{1}{2}sv_{s\ast}^2\mathbf{\hat v_\uparrow}^2\right)^{-2}-1,0\right]}, \label{Eb}
\end{split}
\end{equation}
\end{linenomath*}
where sgn denotes the sign function (note that $\text{sgn}(0)=0$). Note that $\Theta^{\rm roll}=0$ does not imply that a flow with Shields number $\Theta=0$ is able to sustain a rolling motion of grains because $\Theta^{\rm roll}$ depends on the value of $\mathbf{v_\uparrow}$, which is associated with a given periodic saltation trajectory, and for $\Theta=0$, such a trajectory does not exist in the first place.

\subsection{Computation of Threshold and Rate of Equilibrium NST} \label{Computation}
From solving equations~(\ref{uxcomplex}), (\ref{Settling}), (\ref{mub}), (\ref{viz}), (\ref{Thetacompapp}), (\ref{e}), and (\ref{ez}), we obtain a family of identical periodic trajectory solutions $\Theta(Ga,s,S,\hat v_{\uparrow z})$. In detail, for given values of $Ga$, $s$, $S$, and $\hat v_{\uparrow z}$, $\Theta$ is obtained in the following manner:
\begin{enumerate}
 \item Compute $\hat v_{\downarrow z}$ using equation~(\ref{viz}).
 \item Compute $\hat v_{\uparrow x}$ and $\hat v_{\downarrow x}$ using equations~(\ref{e}) and (\ref{ez}).
 \item Compute $v_{s\ast}$ and $\mu_b$ using equation~(\ref{Settling}) and equation~(\ref{mub}), respectively.
 \item Iteratively solve equation~(\ref{Thetacompapp}) for $\Theta$ using equation~(\ref{uxcomplex}). An initial value that works well is $\Theta=1$.
\end{enumerate}
Every periodic trajectory solution $\Theta(Ga,s,S,\hat v_{\uparrow z})$ exhibits a certain value of $sv_{s\ast}^2\mathbf{\hat v_\uparrow}^2$. Hence, $\Theta^{\rm roll}$, calculated as a function of $sv_{s\ast}^2\mathbf{\hat v_\uparrow}^2$ (equation~(\ref{Eb})), can also be parametrized by $Ga$, $s$, $S$, and $\hat v_{\uparrow z}$. Then, from those periodic trajectory solutions that satisfy $\Theta(Ga,s,S,\hat v_{\uparrow z})\geq\Theta^{\rm roll}(Ga,s,S,\hat v_{\uparrow z})$, we obtain the transport threshold from minimizing $\Theta$ as a function of $\hat v_{\uparrow z}$:
\begin{linenomath*}
\begin{equation}
 \Theta_t(Ga,s,S)\equiv\min_{\hat v_{\uparrow z}:\;\Theta(Ga,s,S,\hat v_{\uparrow z})\geq\Theta^{\rm roll}(Ga,s,S,\hat v_{\uparrow z})}\Theta(Ga,s,S,\hat v_{\uparrow z}). \label{Thetat}
\end{equation}
\end{linenomath*}
Note that, for conditions corresponding to the \textit{bedload} regime, associated with low-energy threshold trajectories (a precise definition is provided in section~\ref{TransportRegimes}), periodic trajectory solutions $\Theta(Ga,s,S,\hat v_{\uparrow z})$ tend to be monotonously increasing with $\hat v_{\uparrow z}$, while $\Theta^{\rm roll}(Ga,s,S,\hat v_{\uparrow z})$ is always monotonously decreasing with $\hat v_{\uparrow z}$ (equation~(\ref{Eb})). That is, the threshold trajectory calculated from equation~(\ref{Thetat}) tends to satisfy $\Theta_t=\Theta^{\rm roll}_t\equiv\Theta^{\rm roll}|_{\Theta=\Theta_t}$ for such conditions.

From the threshold trajectory, we obtain the threshold bed friction coefficient $\mu_{bt}$ and dimensionless average streamwise grain velocity $\overline{v_x}_{\ast t}$ using equations~(\ref{mub}), (\ref{LawWall}), (\ref{meanvx}), and (\ref{meanz}):
\begin{linenomath*}
\begin{align}
 \mu_{bt}&=(v_{\downarrow xt}-v_{\uparrow xt})/(v_{\uparrow zt}-v_{\downarrow zt}), \\
 \overline{v_x}_{\ast t}&=\sqrt{\Theta_t}f_u\left[Ga\sqrt{\Theta_t},\frac{1}{2}sv_{s\ast t}^2(\hat v_{\uparrow zt}+\hat v_{\downarrow zt})\right]+(S-\mu_{bt})v_{s\ast t}.
\end{align}
\end{linenomath*}
Lastly, from $\mu_{bt}$, $\overline{v_x}_{\ast t}$, $\Theta_t$, and $c_M=1.7$, we calculate the dimensionless rate $Q_\ast$ of equilibrium NST via equations~(\ref{Q}) and (\ref{M}) using $\mu_{bt}$ as the value of $\mu_b$ in equation~(\ref{M}), since DEM-based numerical simulations of NST indicate that $\mu_b$ does not significantly change with $\Theta$ \cite{PahtzDuran18b}.

\section{Results} \label{Results}
\subsection{Model Evaluation With Experimental and Numerical Data} \label{ModelEvaluation}
\begin{figure}[htb!]
 \begin{center}
  \includegraphics[width=1.0\columnwidth]{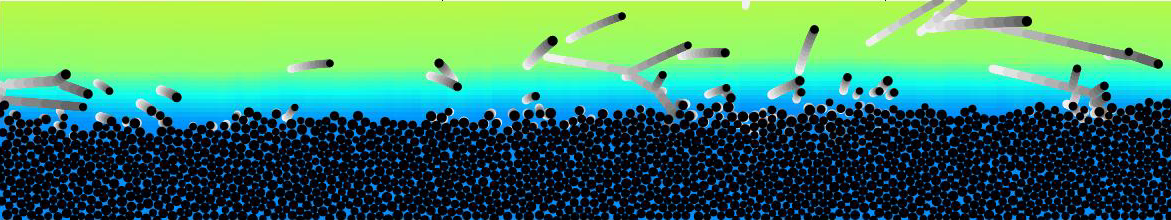}
 \end{center}
 \caption{Snapshot of a simulation of NST using the DEM-based model of \citeA{Duranetal12} showing a quarter of the longitudinal simulation domain. This model couples a discrete element method for the particle motion under gravity, buoyancy, and fluid drag with a continuum Reynolds-averaged description of hydrodynamics. Spherical particles ($\sim 10^4$) with mild polydispersity are confined in a quasi-two-dimensional domain of length $\sim 10^3d$, with periodic boundary conditions in the flow direction, and interact via normal repulsion (restitution coefficient $\epsilon=0.9$) and tangential friction (contact friction coefficient $\mu_c=0.5$). The bottom-most particle layer is glued on a bottom wall, while the top of the simulation domain is reflective but so high that it is never reached by transported particles. The Reynolds-averaged Navier-Stokes equations are combined with a semiempirical mixing length closure that ensures a smooth hydrodynamic transition from high to low particle concentration at the bed surface and quantitatively reproduces the mean turbulent flow velocity profile in the absence of transport. DEM, discrete element method; NST, nonsuspended sediment transport.}
\label{DEM}
\end{figure}

\begin{figure}[htb!]
 \begin{center}
  \includegraphics[width=0.92\columnwidth]{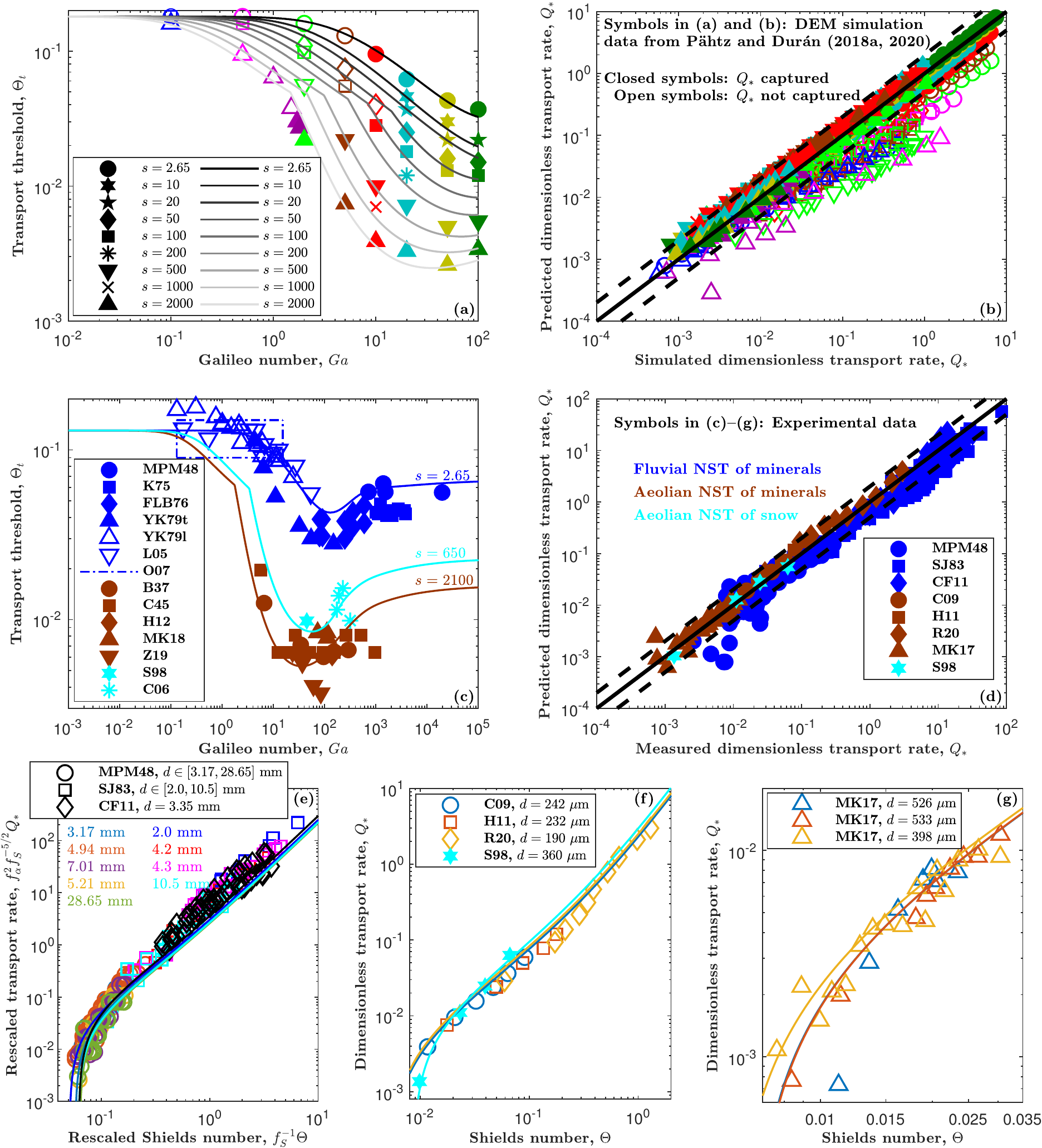}
 \end{center}
 \caption{Model evaluation with experimental and numerical data. (a) and (c) Transport threshold $\Theta_t$ versus Galileo number $Ga$ for varying density ratio $s$. (b) and (d) Predicted versus simulated (b) or measured (d) dimensionless sediment transport rate $Q_\ast$. (e)-(g) Rescaled sediment transport rate $f_\alpha^2f_S^{-5/2}Q_\ast$ versus rescaled Shields number $f_S^{-1}\Theta$, where the slope correction factors $f_\alpha=f_S=1$ in (f) and (g) as $\alpha=0$. Symbols in (a) and (b) correspond to data from DEM-based numerical simulations of NST by \citeA{PahtzDuran18a,PahtzDuran20} for $S=0$ and varying $Ga$ and $s$, as indicated in the legend and plot in (a), and varying $\Theta$ (b). Symbols in (c)-(g) correspond to experimental data of various studies (sections~\ref{FluvialThreshold}-\ref{AeolianRate}). Experimental data by \citeA{Ouriemietal07}, who did not report single data points, lie within the dash-dotted rectangle in (c). Solid lines in (c), (d), (f), and (g) correspond to model predictions. Solid lines in (e) correspond to predictions of the model in which equations~(\ref{Q}) and (\ref{M}) have been approximated by equations~(\ref{Qapp}) and (\ref{Mapp}), respectively. Solid lines in (a) and (b) correspond to model predictions using modified parameter values adjusted to the simulations (section~\ref{ModelEvaluation}). Dashed lines in (b) and (d) indicate a deviation from the predictions by a factor of $2$. DEM, discrete element method; NST, nonsuspended sediment transport.}
\label{Comparison}
\end{figure}

This section compares the model predictions with NST data from many experimental studies and with data from DEM-based numerical simulations of NST by \citeA{PahtzDuran18a};\citeA{PahtzDuran20} using the numerical model of \citeA{Duranetal12} (a snapshot and brief description are provided in Figure~\ref{DEM}).
We start with the comparison to the numerical data in order to explore the range of validity of the model. To this end, the form drag coefficient in equation~(\ref{Settling}) is modified to the value $C^\infty_d=0.5$ used by \citeA{PahtzDuran18a}\citeA{PahtzDuran20}. Furthermore, since \citeA{PahtzDuran18a}\citeA{PahtzDuran20} simulated a quasi-two-dimensional system, parameters characterizing the bed surface also need to be modified. We did so manually and found that the following modified values lead to good overall agreement with the numerical data: $\Theta^o_Y=0.18$, $A=0.8$, $B=0.3$, and $C=0.3$. In fact, Figure~\ref{Comparison}(a) shows that the modified model is consistent with the simulated transport thresholds, calculated via extrapolating $(\mu_b-\tan\alpha)M_\ast$ to zero (consistent with the definition of $\Theta_t$ in equation~(\ref{M})), across a large range of the Galileo number and density ratio: $Ga\in[0.1,100]$ and $s\in[2.65,2000]$.
Furthermore, Figure~\ref{Comparison}(b) shows that the modified model also captures the simulated transport rate data for conditions that satisfy
\begin{linenomath*}
\begin{equation}
 Ga\sqrt{s}\gtrsim
 \begin{cases}
  15,&s<10\\
	75,&s\geq10
 \end{cases}
 \label{ValidityCriterion}
\end{equation}
\end{linenomath*}
within a factor of $2$ (closed symbols), whereas conditions that do not satisfy this constraint are not captured (open symbols). The reason behind this restriction in the model's validity range is that $s^{1/2}Ga$ is a Stokes-like number that encodes the importance of grain inertia relative to viscous drag forcing \cite{PahtzDuran17,PahtzDuran18a}. In fact, the validity of equation~(\ref{Q}) requires that, within the dense, highly collisional part of the transport layer, the motion of a transported grain is not much affected by fluid drag between two subsequent collisions with other transported grains \cite{PahtzDuran20}. Note that collisions between transported grains are not relevant for the transport threshold model, which is why its validity is not affected by $s^{1/2}Ga$ (Figure~\ref{Comparison}(a)).

Figures~\ref{Comparison}(c)-\ref{Comparison}(g) show that the nonmodified model simultaneously captures several experimental data sets corresponding to fluvial and aeolian NST within a factor of about $2$. Note that all the threshold data chosen for the model evaluation were measured using methods that are known or suspected to be consistent with the definition of $\Theta_t$ by equation~(\ref{M}). In the following subsection, the experimental data sets are described in detail.

\subsubsection{Fluvial Transport Threshold Data Sets} \label{FluvialThreshold}
For NST driven by laminar flows, flume measurements of the visual initiation threshold by \citeA<>[YK79l, $s\in(1.9,2.3)$]{YalinKarahan79} and \citeA<>[L05, $s\in(2.1,2.5)$]{Loiseleuxetal05} and of the visual cessation threshold by \citeA<>[O07, $s\in(1.0,2.5)$]{Ouriemietal07} are shown (open symbols and dash-dotted lines in Figure~\ref{Comparison}(c)). Note that, for such conditions, both thresholds are approximately equivalent to each other and $\Theta_t$ \cite{Clarketal17,PahtzDuran18a}. Furthermore, flume measurements of the visual initiation threshold for subaqueous bedload by \citeA<>[K75, $s\simeq2.65$]{Karahan75}, \citeA<>[FLB76, $s\in(1.3,4.6)$]{FernandezLuqueVanBeek76}, and \citeA<>[YK79t, $s\simeq2.65$]{YalinKarahan79} are shown (closed symbols in Figure~\ref{Comparison}(c)), where critical conditions are defined by a critical value of a proxy characterizing the state of transport, such as a critical value of the nondimensionalized bed sediment entrainment rate \cite<>[equation~(7)]{Paphitis01}. It has been argued that the resulting threshold is close to $\Theta_t$ because such and similar critical transport conditions are associated with an abrupt sharp transition in the behavior of $Q_\ast$ as a function of $\Theta$ \cite{Pahtzetal20a}. Furthermore, we obtained $\Theta_t$ from extrapolating the paired flume measurements of $Q_\ast$ and $\Theta$ by \citeA<>[MPM48, those in Figure~\ref{Comparison}(e), $s\in(2.6,2.7)$]{MeyerPeterMuller48} to vanishing $Q_\ast$ using the function $Q_\ast=c_1(\Theta-\Theta_t)+c_2(\Theta-\Theta_t)^2$ \cite<consistent with equations~(\ref{Q}) and (\ref{M}),>[]{PahtzDuran20}, where we treated $c_1$, $c_2$, and $\Theta_t$ as fit parameters.

\subsubsection{Aeolian Transport Threshold Data Sets}
For aeolian NST, we consider data that have been obtained from extrapolating equilibrium transport conditions to vanishing transport, consistent with the definition of $\Theta_t$ in equation~(\ref{M}) as the value at which the equilibrium value of $M_\ast$ vanishes. Following previous studies \cite{Comolaetal19b,MartinKok18,Pahtzetal20a}, we also assume that the cessation threshold of intermittent saltation is identical to $\Theta_t$. However, we do not consider measurements of the continuous transport threshold (discussed in section~\ref{ThresholdInterpretation}) and of the threshold of saltation initiation.

Wind tunnel measurements for windblown sand from \citeA<>[H12, $s\simeq2060$]{Ho12} and \citeA<>[Z19, $s\simeq2210$]{Zhuetal19} are shown in Figure~\ref{Comparison}(c), who carried out an indirect extrapolation to vanishing transport to obtain $\Theta_t$ using a proxy of $Q_\ast$: the surface roughness $z_o$, which undergoes a regime shift when transport ceases. Furthermore, visual wind tunnel measurements for windblown sand and snow by \citeA<>[B37, $s\simeq2210$]{Bagnold37}, \citeA<>[C45, $s\in(1400,1800)$]{Chepil45}, and \citeA<>[S98, $s\simeq650$]{Sugiuraetal98} are shown, who obtained the cessation threshold of intermittent saltation ($\Theta_t$) as the smallest value of $\Theta$ for which energetic mineral or snow grains fed at the tunnel entrance are able to saltate along the bed without stopping. Direct field measurements of this threshold based on the \textit{Time Frequency Equivalence Method} \cite{Wiggsetal04} by \citeA<>[MK18, $s\simeq2210$]{MartinKok18} are also shown. Moreover, wind tunnel measurements of $\Theta_t$ from extrapolating to vanishing transport by \citeA<>[C06, $s\simeq830$]{Cliftonetal06} are shown. However, since \citeA{Cliftonetal06} did not feed snow at the tunnel entrance, we have chosen only their data points for freshly fallen snow. Unlike freshly fallen snow, old snow, used for the other measurements by these authors, is very cohesive \cite{PomeroyGray90}, and NST of old snow therefore requires a distance to reach equilibrium that is very likely much longer than the length of the wind tunnel of \citeA{Cliftonetal06} in the absence of snow feeding \cite{Comolaetal19a}. However, we have not excluded cohesive measurements if sediment feeding occurred, such as the two windblown sand data points at $Ga\approx5$ by \citeA{Bagnold37} and \citeA{Chepil45}, corresponding to small and thus cohesive mineral grains with $d\approx75~\mu\mathrm{m}$, and the windblown snow data point by \citeA{Sugiuraetal98}, corresponding to potentially very cohesive old snow.

\subsubsection{Fluvial Transport Rate Data Sets}
Since the model does not capture $Q_\ast$ for conditions for which $s^{1/2}Ga$ is too small (Figure~\ref{Comparison}(b)), we compare it only to subaqueous bedload measurements of $Q_\ast$, for which $s^{1/2}Ga$ is sufficiently large. In Figures~\ref{Comparison}(d) and \ref{Comparison}(e), the flume measurements of $Q_\ast$ by \citeA<>[MPM48, $s\in(2.6,2.7)$]{MeyerPeterMuller48}, as corrected by \citeA{WongParker06}, for relatively weak driving flows and small slope numbers ($S\simeq0$) and by \citeA<>[SJ83, $s\in(2.6,2.7)$]{SmartJaeggi83} and \citeA<>[CF11, $s\simeq1.51$]{CapartFraccarollo11} for relatively intense driving flows and large $S$ are shown. For all these data sets, the applied fluid shear stress is defined as $\tau=\rho_fgh_m\sin\alpha$ (\ref{MomentumBalances}), where $h_m$ is the flow depth, including the sediment-fluid mixture above the bed surface $z=0$, and we corrected $h_m$ for side wall drag using the method described in section~2.3 of \citeA{Guo14}. For the latter two data sets, the model predictions can depend significantly on $\alpha$ and $S$ for a given $\Theta$. In order to make $Q_\ast$ only dependent on a single rather than three independent external control parameters, we have approximated equations~(\ref{Q}) and (\ref{M}) in Figure~\ref{Comparison}(e) (but not in Figure~\ref{Comparison}(d)). We have used (\ref{AppendixNoVerticalDrag}) $\mu_b\approx\mu^o_b\equiv\lim_{e_z\rightarrow1}\mu_b\simeq0.646$, $\Theta_t\approx f_S\Theta_t|_{S=0}$, and $\overline{v_x}_{\ast t}\approx f_S^{1/2}\overline{v_x}_{\ast t}|_{S=0}$, where $f_S\equiv1-S/\mu^o_b$, and we have used $f_\alpha f_S^{-1}(1+c_MM_\ast)\approx1+c_Mf_\alpha f_S^{-1}M_\ast$ because the terms $f_\alpha\equiv1-\tan\alpha/\mu^o_b$ and $f_S$ predominantly matter when $\alpha$ (and thus $M_\ast$) is large. Combined, these approximations yield
\begin{subequations}
\begin{linenomath*}
\begin{align}
 f_\alpha^2f_S^{-5/2}Q_\ast&\approx f_\alpha f_S^{-1}M_\ast\overline{v_x}_{\ast t}|_{S=0}(1+c_Mf_\alpha f_S^{-1}M_\ast), \label{Qapp} \\
 f_\alpha f_S^{-1}M_\ast&\approx(f_S^{-1}\Theta-\Theta_t|_{S=0})/\mu^o_b, \label{Mapp}
\end{align}
\end{linenomath*}
\end{subequations}
which are expressions independent of $\alpha$ and $S$ for the rescaled transport rate $f_\alpha^2f_S^{-5/2}Q_\ast$ as a function of the rescaled Shields number $f_S^{-1}\Theta$. Note that $f_S\simeq f_\alpha$ for aeolian NST and slope-driven NST in viscous liquids, implying that this slope correction of $\Theta$ and $Q_\ast$ is consistent with previous results for windblown sand \cite{IversenRasmussen99,Wangetal21}, while $f_S<f_\alpha$ for slope-driven NST in turbulent liquids. These differences reflect the different physical origins of $f_S$ and $f_\alpha$. The physics behind $f_S$ involves the streamwise buoyancy force acting on transported grains, which is associated with the gradient of the viscous fluid shear stress \cite{Maurinetal18}, while the physics behind $f_\alpha$ involves the average streamwise fluid momentum balance (\ref{MomentumBalances}), which is associated with the gradient of the sum of the viscous and Reynolds-averaged fluid shear stress.

\subsubsection{Aeolian Transport Rate Data Sets} \label{AeolianRate}
For windblown sand and snow, wind tunnel measurements of $Q_\ast$ by \citeA<>[C09, $s\simeq2080$]{Creysselsetal09}, \citeA<>[H11, $s\simeq2060$]{Hoetal11}, \citeA<>[R20, $s\simeq2210$]{Ralaiarisoaetal20}, and \citeA<>[S98, $s\simeq650$]{Sugiuraetal98} are shown in Figures~\ref{Comparison}(d) and \ref{Comparison}(f). Note that the experiments by \citeA{Sugiuraetal98} were carried out using potentially very cohesive old snow, while the data set by \citeA{Ralaiarisoaetal20} corresponds to the first controlled measurements of $Q_\ast$ for intense windblown sand, which are not captured by standard expressions for $Q_\ast$ from the literature. Furthermore, field measurements of $Q_\ast$ for windblown sand by \citeA<>[MK17, $s\simeq2210$]{MartinKok17} are shown in Figures~\ref{Comparison}(d) and \ref{Comparison}(g). These authors estimated the intermittent (i.e., nonequilibrium) transport rate $Q^{\rm in}$ and the fraction $f_Q$ of active windblown sand, from which we obtained the equilibrium transport rate via $Q=Q^{\rm in}/f_Q$ \cite{Comolaetal19b}.

\subsection{NST Regimes} \label{TransportRegimes}
\begin{figure}[htb!]
 \begin{center}
  \includegraphics[width=1.0\columnwidth]{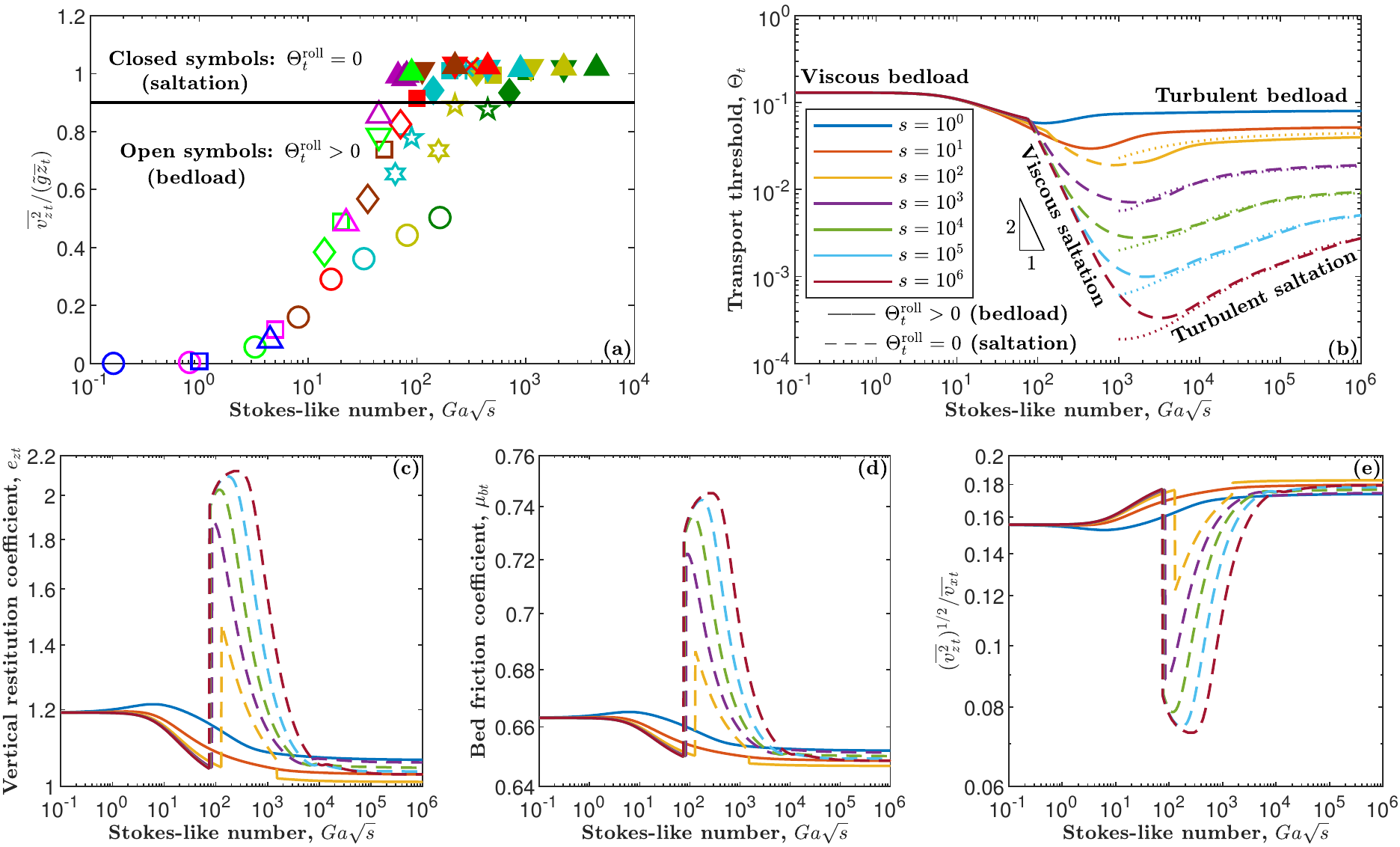}
 \end{center}
 \caption{Transport regimes for threshold conditions. (a) Fraction $\overline{v_z^2}_t/(\tilde g\overline{z}_t)$ of transport layer thickness due to the contact-free motion of grains, (b) transport threshold $\Theta_t$, (c) vertical restitution coefficient $e_{zt}$, (d) bed friction coefficient $\mu_{bt}$, and (e) $(\overline{v_z^2}_t)^{1/2}/\overline{v_x}_t$ versus Stokes-like number $s^{1/2}Ga$. Symbols in (a) correspond to data from DEM-based numerical simulations of NST by \citeA{PahtzDuran18a};\citeA{PahtzDuran20} for $S=0$ and varying $Ga$ and $s$, where the symbol openness indicates the value of $\Theta^{\rm roll}_t$ predicted by the transport threshold model using modified parameter values adjusted to the simulations (section~\ref{ModelEvaluation}): $\Theta^{\rm roll}_t>0$ (i.e., bedload) for open symbols and $\Theta^{\rm roll}_t=0$ (i.e., saltation) for closed symbols. Symbol shapes and colors in (a) are the same as in Figure~\ref{Comparison}(a). The solid line in (a) indicates $\overline{v_z^2}_t/(\tilde g\overline{z}_t)=0.9$. Solid and dashed lines in (b)-(e) indicate model predictions for $S=0$ using the original parameter values. Solid lines indicate a predicted value $\Theta^{\rm roll}_t>0$ and dashed lines the predicted value $\Theta^{\rm roll}_t=0$. Dotted lines in (b) correspond to the model approximation for turbulent saltation (equation~(\ref{ThetatTurbulentSaltation2})), valid for $s^{1/4}Ga\gtrsim200$. Figure legends in (c)-(e) are the same as in (b). DEM, discrete element method; NST, nonsuspended sediment transport.}
\label{ModelApproximations}
\end{figure}

\citeA{PahtzDuran18a} provided a criterion to distinguish \textit{bedload}, defined as NST in which a significant portion of transported grains is moving in enduring contacts with the bed surface (e.g., via rolling and sliding), from \textit{saltation}, defined as NST in which this portion is insignificant. This criterion states that saltation occurs when more than $90\%$ of the transport layer thickness are due to the contact-free motion of grains: $\overline{v_z^2}/(\tilde g\overline{z})\geq0.9$. Here, for threshold conditions, we distinguish bedload from saltation using the threshold trajectory value of the critical Shields number required for rolling $\Theta^{\rm roll}_t$, which is associated with the rebound energy of the threshold trajectory. When $\Theta^{\rm roll}_t=0$, grains are able to escape the pockets of the bed surface solely due to their saltation motion, that is, without the assistance of the near-surface flow. Hence, we identify this regime as saltation and distinguish it from bedload where $\Theta^{\rm roll}_t>0$. Figure~\ref{ModelApproximations}(a) shows that this criterion is consistent with the one by \citeA{PahtzDuran18a} for these authors' data obtained from their DEM-based simulations of NST.

Based on the hop height calculated from the transport threshold model, $H_t=[v_{\uparrow zt}v_{st}-v_{st}^2\ln(1+v_{\uparrow zt}/v_{st})]/\tilde g$ (equation~\ref{H}), and the thickness of the viscous sublayer of the turbulent boundary layer, $\delta_{\nu t}=10d/Re_{dt}$, we distinguish between \textit{viscous} ($H_t\lesssim\delta_{\nu t}$) and \textit{turbulent} ($H_t\gtrsim\delta_{\nu t}$) conditions, giving rise to totally four transport regimes, which are indicated by text in Figure~\ref{ModelApproximations}(b): \textit{viscous bedload}, \textit{turbulent bedload}, \textit{viscous saltation}, and \textit{turbulent saltation}. It can be seen that viscous bedload occurs when the Stokes-like number $s^{1/2}Ga$ falls below about $75$, implying that the validity criterion for the model's transport rate predictions (equation~(\ref{ValidityCriterion})) is only disobeyed for viscous bedload conditions. Note that the transition from viscous bedload to viscous saltation coincides with a kink in the threshold curves (Figures~\ref{Comparison}(a), \ref{Comparison}(c), and \ref{ModelApproximations}(b)) and that turbulent saltation occurs when $s^{1/4}Ga\gtrsim200$ (section~\ref{SlopeDependency}).

Figure~\ref{ModelApproximations}(b) shows that, for viscous saltation, the transport threshold approximately scales as $\Theta_t\propto(s^{1/2}Ga)^{-2}$. To demonstrate the origin of this scaling, we approximate the flow velocity profile in equation~(\ref{uxcomplex}) as $u_x\simeq u_\ast Re_dz/d$, since saltation trajectories are relatively large ($H\gg Z_\Delta d$) and fully submerged within the viscous sublayer. Using this profile in equation~(\ref{Thetacompapp}) and approximating the dimensionless terminal settling velocity by its Stokes drag limit, $v_{s\ast}\simeq Ga/18$ (equation~(\ref{Settling}) for small $Ga$), yields
\begin{subequations}
\begin{linenomath*}
\begin{alignat}{2}
 \text{With vertical drag:}\quad&&\Theta&=18\frac{\mu_b(1+\hat v_{\uparrow z})+\hat v_{\uparrow x}-S}{-\hat v_{\downarrow z}(1+\hat v_{\uparrow z})-\hat v_{\uparrow z}}\left(Ga\sqrt{s}\right)^{-2}, \label{ViscousSaltation} \\
 \text{Neglected vertical drag:}\quad&&\Theta&\simeq18\frac{\mu^o_b(1+\hat v_{\uparrow z})+\hat v_{\uparrow x}-S}{\frac{1}{3}\hat v_{\uparrow z}^2}\left(Ga\sqrt{s}\right)^{-2}, \label{ViscousSaltationapp}
\end{alignat}
\end{linenomath*}
\end{subequations}
where equation~(\ref{ViscousSaltationapp}) is the approximation of equation~(\ref{ViscousSaltation}) valid for negligible vertical drag (\ref{AppendixNoVerticalDrag}). After linking $\hat v_{\downarrow z}$, $\hat v_{\uparrow x}$, $\hat v_{\downarrow x}$, and $\mu_b$ to $\hat v_{\uparrow z}$ via equations~(\ref{viz}), (\ref{e}), and (\ref{ez}), the crucial difference between both equations is that $\Theta$ in equation~(\ref{ViscousSaltation}) first monotonously decreases with $\hat v_{\uparrow z}$ until it approaches a minimum and then monotonously increases with $\hat v_{\uparrow z}$, whereas $\Theta$ in equation~(\ref{ViscousSaltationapp}) monotonously decreases with $\hat v_{\uparrow z}$ for the entire range of $\hat v_{\uparrow z}$. Hence, obtaining the transport threshold $\Theta_t$ from equation~(\ref{ViscousSaltation}) via minimizing $\Theta$ (equation~(\ref{Thetat})) yields $\Theta_t\propto(s^{1/2}Ga)^{-2}$ for a fixed $S$, whereas the use of equation~(\ref{ViscousSaltationapp}) would yield a contradiction: an infinitely large threshold trajectory ($\hat v_{\uparrow zt}=\infty$) and $\Theta_t=0$. Hence, vertical drag is not negligible in viscous saltation. In fact, the spikes in Figure~\ref{ModelApproximations}(c) indicate a substantial positive deviation from $e_z=1$ for viscous saltation, whereas $e_z$ is close to unity for the other regimes. Larger values of $e_z$ mean that vertical drag suppresses the vertical motion of grains more strongly, causing a slight positive deviation from $\mu_{bt}=\mathrm{const}$ (spikes in Figure~\ref{ModelApproximations}(d)) and a substantial negative deviation from $(\overline{v_z^2}_t)^{1/2}\propto\overline{v_x}_t$ (spikes in Figure~\ref{ModelApproximations}(e)). In \ref{SaltationJustification}, we use these two scaling relations and their violation for viscous saltation to justify the rebound boundary conditions.

We now perform a similar analysis for turbulent saltation. We approximate the flow velocity profile in equation~(\ref{uxcomplex}) as the log-layer profile $u_x\simeq\kappa^{-1}u_\ast\ln(z/z_o)$, where $z_o$ is the roughness height, since saltation trajectories are relatively large ($H\gg Z_\Delta d$) and substantially exceed the viscous sublayer and buffer layer. Using this profile in equation~(\ref{Thetacompapp}) and neglecting vertical drag (as $e_z\simeq1$, see Figure~\ref{ModelApproximations}(c)) yields
\begin{linenomath*}
\begin{equation}
 \frac{\sqrt{\Theta}}{\kappa v^o_{s\ast}}\ln\left(\frac{sv_{s\ast}^{o2}\hat v_{\uparrow z}^2}{3z_o/d}\right)\simeq\mu^o_b(1+\hat v_{\uparrow z})+\hat v_{\uparrow x}-S, \label{TurbulentSaltationapp}
\end{equation}
\end{linenomath*}
where $v^o_{s\ast}\equiv\lim_{e_z\rightarrow1}v_{s\ast}=v_{s\ast}|_{\mu_b=\mu^o_b}$. Using that $\hat v_{\uparrow x}\propto\hat v_{\uparrow z}$ for conditions in which vertical drag is negligible (\ref{AppendixNoVerticalDrag}), we can minimize $\Theta$ in equation~(\ref{TurbulentSaltationapp}) via $\mathrm{d}\Theta/\mathrm{d}\hat v_{\uparrow z}=0$ to obtain $\Theta_t$ (equation~(\ref{Thetat})). This yields after some rearrangement $\mu^o_b\hat v_{\uparrow zt}+\hat v_{\uparrow xt}=2\sqrt{\Theta_t}/(\kappa v^o_{s\ast})$ and subsequently, using the definition of the hat,
\begin{linenomath*}
\begin{equation}
 \frac{\sqrt{\Theta_t}}{\kappa}\left[\ln\left(\frac{\beta s\Theta_t}{\kappa^2z_{ot}/d}\right)-2\right]\simeq(\mu^o_b-S)v^o_{s\ast}, \label{ThetatTurbulentSaltation}
\end{equation}
\end{linenomath*}
where $\beta\equiv\lim_{e_z\rightarrow1}(4v_{\uparrow z}^2/3)/(\mu_bv_{\uparrow z}+v_{\uparrow x})^2\simeq0.136$ (\ref{AppendixTurbulentSaltation}). For both $z_{ot}=d/30$, valid in the fully rough regime ($Re_d\gtrsim70$), and $z_{ot}=d/(9Re_{dt})$, valid in the fully smooth regime ($Re_d\lesssim4$), equation~(\ref{ThetatTurbulentSaltation}) can be explicitly solved (\ref{AppendixTurbulentSaltation}). The maximum of both solutions (dotted lines in Figure~\ref{ModelApproximations}(b)) well approximates $\Theta_t$ for turbulent saltation:
\begin{linenomath*}
\begin{equation}
\begin{split}
 \Theta_t&\simeq\max(\Theta^{\rm rough}_t,\Theta^{\rm smooth}_t),\text{ with} \\
 \Theta^{\rm rough}_t&\equiv\left(\frac{\kappa(\mu^o_b-S)v^o_{s\ast}}{2W\left[(\mu^o_b-S)v^o_{s\ast}\sqrt{15e^{-2}\beta s/2}\right]}\right)^2\text{ and} \\
 \Theta^{\rm smooth}_t&\equiv\left(\frac{\kappa(\mu^o_b-S)v^o_{s\ast}}{3W\left[(\mu^o_b-S)v^o_{s\ast}\sqrt[3]{e^{-2}\kappa\beta sGa/3}\right]}\right)^2. \label{ThetatTurbulentSaltation2}
\end{split}
\end{equation}
\end{linenomath*}

\subsection{Bed Slope Dependency of Transport Threshold} \label{SlopeDependency}
\begin{figure}[htb!]
 \begin{center}
  \includegraphics[width=1.0\columnwidth]{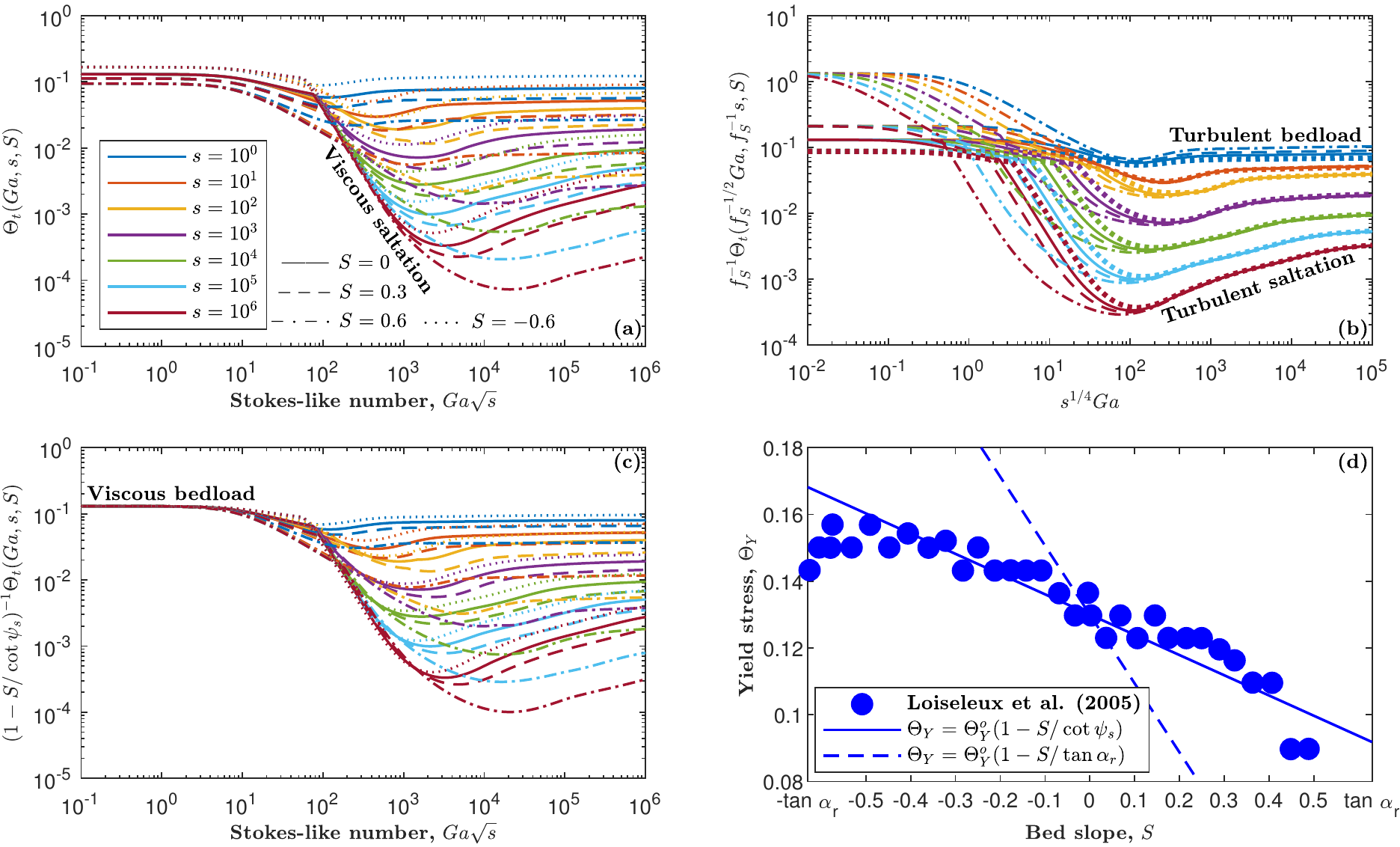}
 \end{center}
 \caption{Effect of slope number $S$ on transport threshold predictions. (a) Transport threshold $\Theta(Ga,s,S)$ versus Stokes-like number $s^{1/2}Ga$. (b) Rescaled transport threshold $f_S^{-1}\Theta_t(f_S^{-1/2}Ga,f_S^{-1}s,S)$, where $f_S=1-\mu^o_b/S$, versus the dimensionless number $s^{1/4}Ga$, which roughly controls the occurrence of the transport threshold minimum for saltation \cite{PahtzDuran18a}. Note that, according to equation~(\ref{STurbulentNST}), $f_S^{-1}\Theta_t(f_S^{-1/2}Ga,f_S^{-1}s,S)$ is approximately equal to $\Theta_t(Ga,s,0)$ (independent of $S$) for turbulent NST. (c) Rescaled transport threshold $(1-S/\cot\psi_s)^{-1}\Theta_t(Ga,s,S)$ versus $s^{1/2}Ga$. (d) Yield stress $\Theta_Y$ versus $S$. Lines in (a)-(c) correspond to model predictions. Symbols in (d) correspond to measurements of $\Theta_t$ by \citeA{Loiseleuxetal05} for $Ga=[0.17,0.54]$ and $s=2.5$ (i.e., $\Theta_t\simeq\Theta_Y$) and various $S$. Measurements of the angle of repose $\alpha_r$ by \citeA{Loiseleuxetal05} are indicated in the abscissa of (d). The solid line in (d) corresponds to the slope correction predicted by the model for viscous bedload (equation~(\ref{SViscousBedload})). The dashed line in (d) corresponds to the classical slope correction. Figure legends in (b) and (c) are the same as in (a). NST, nonsuspended sediment transport.}
\label{Slope}
\end{figure}

Figure~\ref{Slope} shows how the slope number $S$ affects the transport threshold predictions.
For the different NST regimes, different scaling laws are found:
\begin{subequations}
\begin{linenomath*}
\begin{alignat}{2}
 \text{Viscous saltation:}\quad&&\Theta_t(Ga,s,S)&\simeq\Theta_t(Ga,s,0), \label{SViscousSaltation} \\
 \text{Turbulent NST:}\quad&&\Theta_t(Ga,s,S)&\simeq f_S\Theta_t(\sqrt{f_S}Ga,f_Ss,0),\quad\text{and} \label{STurbulentNST} \\
 \text{Viscous bedload:}\quad&&\Theta_t(Ga,s,S)&\simeq\Theta_Y=(1-S/\cot\psi_s)\Theta^o_Y, \label{SViscousBedload}
\end{alignat}
\end{linenomath*}
\end{subequations}
the validity of which is shown in Figures~\ref{Slope}(a), \ref{Slope}(b), and \ref{Slope}(c), respectively. Equation~(\ref{SViscousSaltation}) follows from the fact that the term $\mu_b(1+\hat v_{\uparrow z})+\hat v_{\uparrow z}$ in equation~(\ref{ViscousSaltation}) is substantially larger than $S$ for the threshold trajectory, since grain velocities become comparable to the terminal settling velocity $v_s$ (i.e., $\hat v_{\uparrow zt}\sim1$) because of vertical drag in viscous saltation, implying that the effect of $S$ is small. In contrast, for turbulent NST, vertical drag can be neglected (i.e., $\hat v_{\uparrow zt}\ll1$), which leads to equation~(\ref{STurbulentNST}) (\ref{AppendixNoVerticalDrag}). Note that, for turbulent saltation, the validity of equation~(\ref{STurbulentNST}) requires $s^{1/4}Ga\gtrsim200$, where $s^{1/4}Ga$ is a dimensionless number that roughly controls the occurrence of the transport threshold minimum for saltation \cite{PahtzDuran18a}, consistent with Figure~\ref{Slope}(b). Equation~(\ref{SViscousBedload}) follows from equation~(\ref{Eb}) as, for viscous bedload, $\frac{1}{2}sv_{s\ast}^2\mathbf{\hat v_\uparrow}^2\propto Ga^2s$ becomes negligible (section~\ref{ReboundEnergy}).

Classically, the term $1-S/\tan\alpha_r$, where $\alpha_r$ is the angle of repose (i.e., the slope angle at which the entire granular bed, as opposed to only the bed surface, fails), is used to correct $\Theta$ and/or $\Theta_t$ in NST \cite{IversenRasmussen94,Maurinetal18}, in reasonable agreement with threshold measurements \cite{ChiewParker94,IversenRasmussen94}. Both the slope correction term $f_S\equiv1-S/\mu^o_b$ in equation~(\ref{STurbulentNST}) and the slope correction term $1-S/\cot\psi_s$ in equation~(\ref{SViscousBedload}) resemble the functional structure of $1-S/\tan\alpha_r$. However, $\mu^o_b$ is a purely kinematic quantity and entirely unrelated to $\tan\alpha_r$ even though its value ($\mu^o_b\simeq0.646$) is very close to typical values of $\tan\alpha_r$. Likewise, $\cot\psi_s\simeq2.14$ is substantially larger than typical values of $\tan\alpha_r$ and the predicted slope correction for viscous bedload therefore much milder than the classical one. Consistently, Figure~\ref{Slope}(d) shows that, for the viscous bedoad experiments by \citeA{Loiseleuxetal05}, the model (solid line) reproduces the measured behavior that $\Theta_t$ changes only mildly with $S$ for $|S|\lesssim0.5$, while the classical slope correction (dashed line) fails to capture these data. The deviations between model and measurements for $|S|\gtrsim0.5$ are likely due to the fact that $|S|$ approaches $\tan\alpha_r$, weakening the resistance of the bulk of the bed. In fact, once the bulk of the bed is close to yield, this will probably affect the resistance of bed surface grains via long-range correlations \cite<since yielding is probably a critical phenomenon,>[]{Pahtzetal20a}, which the model does not account for.

Lastly, we emphasize that the model predictions do not take into account that large bed slopes in nature (e.g., for mountain streams) are usually accompanied by very small flow depths of the order of $1d$, which cause bed mobility to decrease rather than increase with $S$~\cite{PrancevicLamb15,Prancevicetal14}.

\section{Discussion} \label{Discussion}
\subsection{Transport Threshold Interpretation} \label{ThresholdInterpretation}
In section~\ref{Model}, we first idealized equilibrium NST in a manner that eliminates all kinds of grain trajectory fluctuations. We then defined the transport threshold $\Theta_t$ as the smallest Shields number for which a nontrivial steady state grain trajectory exists. This definition raises two important questions: 
\begin{itemize}
 \item What values of the initial lift-off velocity $\mathbf{v^o_\uparrow}$ are sufficient for a grain to approach the nontrivial steady state threshold trajectory?
 \item How sensitive is this threshold trajectory to trajectory fluctuations?
\end{itemize}
This section gives answers to these questions and discusses consequences for the physical interpretation of $\Theta_t$ arising from these answers.

We reiterate that, for most bedload conditions, the model predicts $\Theta_t=\Theta^{\rm roll}_t$ (section~\ref{Computation}). This implies that the average flow is barely able to sustain a rolling motion of grains out off the most stable pockets of the bed surface, meaning that grain motion would stop for an arbitrarily small negative fluctuation of the threshold rebound velocity $\mathbf{v_{\uparrow t}}$, since smaller grain energies make rolling out of the most stable bed surface pockets more difficult (section~\ref{ReboundEnergy}).

\begin{figure}[htb!]
 \begin{center}
  \includegraphics[width=0.97\columnwidth]{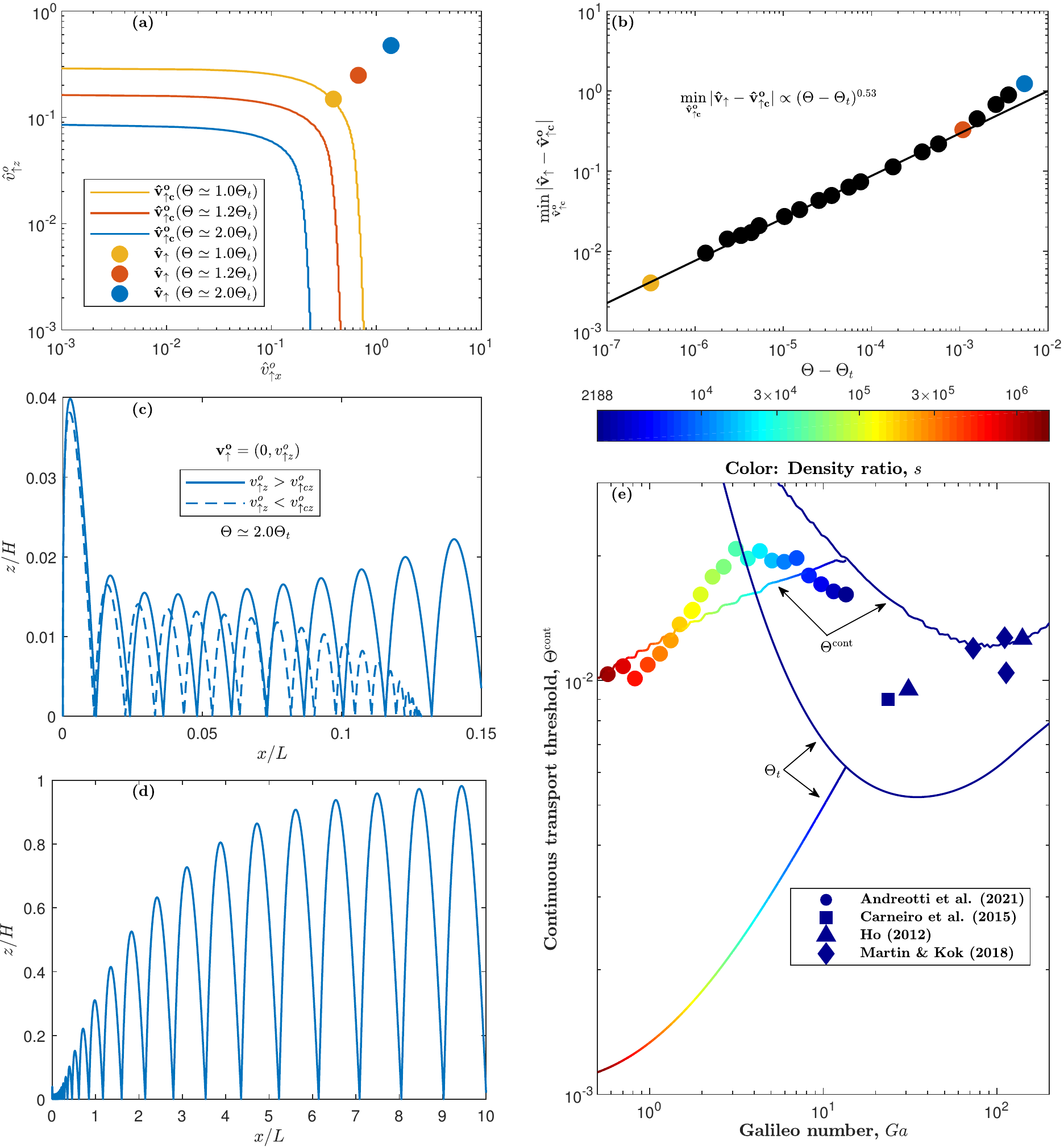}
 \end{center}
 \caption{Grain trajectories and continuous transport threshold $\Theta^{\rm cont}$. (a) Critical lines $\mathbf{\hat v^o_{\uparrow c}}$ separating supercritical conditions (northeast of the lines) from subcritical conditions (southwest of the lines) exemplary for the case $s=2000$, $Ga=40$, $S=0$, and three different Shields numbers $\Theta$. Supercritical refers to initial conditions ($\mathbf{\hat v^o_\uparrow}$) that approach the periodic saltation trajectory solution $\mathbf{\hat v_\uparrow}$ (circles). Subcritical refers to initial conditions that approach no motion. (b) Critical scaling of the distance $\min_{\mathbf{\hat v^o_{\uparrow c}}}|\mathbf{\hat v_\uparrow}-\mathbf{\hat v^o_{\uparrow c}}|$ between $\mathbf{\hat v_\uparrow}$ and $\mathbf{\hat v^o_{\uparrow c}}$. (c) Examples for grain trajectories $z(x)$, where $z$ and $x$ are normalized by the hop height $H$ and hop length $L$, respectively, of the periodic saltation trajectory, with supercritical (solid lines) and subcritical (dashed lines) initial conditions $\mathbf{v^o_\uparrow}=(0,v^o_{\uparrow z})$ for $\Theta\simeq2.0\Theta_t$. (d) The same plot as in (c) but for a much wider range of $x/L$. (e) $\Theta^{\rm cont}$ versus Galileo number $Ga$. Symbols in (e) correspond to windblown sand measurements of $\Theta^{\rm cont}$ for either $d=125~\mu\mathrm{m}$ and varying $\rho_f$ \cite{Andreottietal21} or $\rho_f=1.2~\mathrm{kg/m}^3$ and varying $d$ \cite{Carneiroetal15,Ho12,MartinKok18}. The symbol color indicates the value of the density ratio $s$. Lines in (e) correspond to model predictions of $\Theta^{\rm cont}$ and the transport threshold $\Theta_t$.}
\label{Stability}
\end{figure}

For saltation, the model predicts a similar behavior as for bedload. To show this, we calculate the saltation trajectory evolution from equations~(\ref{zt}), (\ref{vzt}), (\ref{xtapp}), and (\ref{vxtapp}) using equation~(\ref{uxcomplex}) and the periodic saltation trajectory (i.e., steady) value of the dimensionless terminal settling velocity $v_{s\ast}$, the calculation of which was described in section~\ref{Computation}. Based on this calculation, Figure~\ref{Stability}(a) shows for an exemplary saltation case ($s=2000$, $Ga=40$, $S=0$, and $\Theta\simeq[1.0,1.2,2.0]\Theta_t$) the critical lines $\mathbf{\hat v^o_{\uparrow c}}$ separating those conditions with an initial dimensionless lift-off velocity $\mathbf{\hat v^o_\uparrow}$ that approach the periodic trajectory solution $\mathbf{\hat v_\uparrow}$ (supercritical, northeast of the critical lines, see solid lines in Figures~\ref{Stability}(c) and \ref{Stability}(d) for an exemplary trajectory) from those conditions that approach no motion (subcritical, southwest of the critical lines, see dashed line in Figure~\ref{Stability}(c) for an exemplary trajectory).
Furthermore, Figure~\ref{Stability}(b) shows that the distance $\min_{\mathbf{\hat v^o_{\uparrow c}}}|\mathbf{\hat v_\uparrow}-\mathbf{\hat v^o_{\uparrow c}}|$ between $\mathbf{\hat v_\uparrow}$ and $\mathbf{\hat v^o_{\uparrow c}}$ obeys critical scaling behavior for sufficiently small $\Theta-\Theta_t$ and vanishes in the limit $\Theta\rightarrow\Theta_t$. Hence, arbitrarily small negative fluctuations of $\mathbf{v_{\uparrow t}}$ will cause saltation to cease in time.

The model prediction that both bedload and saltation threshold trajectories are unstable against arbitrarily small negative fluctuations of $\mathbf{v_{\uparrow t}}$ implies that, for realistic natural settings associated with substantial grain trajectory fluctuations, continuous NST cannot be sustained once $\Theta$ is too close to $\Theta_t$. In other words, $\Theta_t$ must be strictly smaller than the continuous transport threshold $\Theta^{\rm cont}$ regardless of how continuous NST is defined \cite<an unambiguous definition is currently an open problem,>[section~4.1.3.2]{Pahtzetal20a}.

This subsection is subdivided into further subsections. Section~\ref{ContinuousTransport} models $\Theta^{\rm cont}$ for aeolian NST and compares the model predictions with experimental data. Section~\ref{IntermittentTransport} discusses intermittent NST dynamics occurring for $\Theta<\Theta^{\rm cont}$.

\subsubsection{Continuous Transport Threshold} \label{ContinuousTransport}
Understanding the behavior of $\Theta^{\rm cont}$ is crucial because equilibrium transport rate expressions, such as equations~(\ref{Q}) and (\ref{M}), are invalid for intermittent (i.e., nonequilibrium) NST conditions \cite{Comolaetal19b,Pahtzetal20a}. To this end, let us consider NST conditions with $\Theta>\Theta_t$ and suppose that the system departs more and more from the equilibrium by depositing grains on the bed surface. The more grains are deposited, the more the flow will be undisturbed by the presence of transported grains. To drive this system back to equilibrium, it is required that bed surface grains are entrained and subsequently net accelerated by the undisturbed flow. We therefore propose that $\Theta^{\rm cont}$ corresponds to the minimal Shields number for which the average velocity of such an entrained grain $\mathbf{v^e_\uparrow}$ is supercritical, denoted by $\mathbf{v^e_\uparrow}\succeq\mathbf{v^o_{\uparrow c}}$ (note that both $\mathbf{v^e_\uparrow}$ and $\mathbf{v^o_{\uparrow c}}$ depend on the undisturbed flow condition $(Ga,s,S,\Theta)$):
\begin{linenomath*}
\begin{equation}
 \Theta^{\rm cont}(Ga,s,S)\equiv\min_{\hat v_{\uparrow z}:\;\mathbf{v^e_\uparrow}(Ga,s,S,\Theta)\succeq\mathbf{v^o_{\uparrow c}}(Ga,s,S,\Theta)}\Theta(Ga,s,S,\hat v_{\uparrow z}), \label{Thetacont}
\end{equation}
\end{linenomath*}
where $\Theta(Ga,s,S,\hat v_{\uparrow z})$ denotes, like before (section~\ref{Computation}), a periodic trajectory solution. Consistent with this definition, Figure~\ref{Stability}(a) shows that, for saltation, the range of supercritical initial dimensionless lift-off velocities ($\mathbf{\hat v^o_\uparrow}\succeq\mathbf{\hat v^o_{\uparrow c}}$) substantially increases with $\Theta/\Theta_t$. Note that our proposed definition of $\Theta^{\rm cont}$ is similar to the one by \citeA{DoorschotLehning02}. The main and possibly only difference is that these authors' definition referred to the average grain lifting off from the bed surface, including entrained and rebounding grains, rather than only the average entrained grain.

For aeolian NST, we can model $\Theta^{\rm cont}$ using equation~(\ref{Thetacont}) and the fact that, on average, bed surface grains entrained by impacts of transported grains are more energetic than grains entrained directly by the flow and therefore more likely to be supercritical. In fact, the former grains are literally ejected and their average ejection velocity $\mathbf{v^e_\uparrow}$ is weakly but significantly correlated with the impact velocity $\mathbf{v_\downarrow}$ \cite{Beladjineetal07}. An empirical scaling relation that is consistent with experimental data is
\begin{linenomath*}
\begin{subequations}
\begin{align}
  v^e_{\uparrow x}&=0, \label{vex} \\
  v^e_{\uparrow z}/\sqrt{\tilde gd}&=c_e(|\mathbf{v_\downarrow}|/\sqrt{\tilde gd})^{1/4}, \label{vez}
\end{align}
\end{subequations}
\end{linenomath*}
where $c_e$ is a proportionality constant close to unity \cite{Beladjineetal07}. Figure~\ref{Stability}(e) shows model predictions of $\Theta^{\rm cont}$, using equations~(\ref{Thetacont}), (\ref{vex}), and (\ref{vez}) with $c_e=0.8$, and compares them with measurements of $\Theta^{\rm cont}$ for aeolian NST \cite{Andreottietal21,Carneiroetal15,Ho12,MartinKok18}. Note that \citeA{Andreottietal21} carried out their experiments in a wind tunnel with varying air pressure and defined threshold conditions as the transition between saltation of groups of particles (bursts) to intermittent saltation of single particles (at high pressure) or no transport (at low pressure). These authors interpreted their measurements as measurements of the transport threshold $\Theta_t$, that is, they assumed $\Theta^{\rm cont}=\Theta_t$. However, Figure~\ref{Stability}(e) shows that, for many of their measured conditions, the model predictions of $\Theta_t$ are nearly an order of magnitude below those of $\Theta^{\rm cont}$, and only the latter are consistent with the measurements.

\subsubsection{Intermittent NST} \label{IntermittentTransport}
For Shields numbers $\Theta$ below $\Theta^{\rm cont}$, NST may remain intermittent. There are two distinct kinds of transport intermittency. The first kind occurs when turbulence-driven bed sediment entrainment events associated with energetic turbulent eddies \cite{Cameronetal20,Paternaetal16,Valyrakisetal11,Zhengetal20} generate intermittent rolling events of entrained grains \cite{Pahtzetal20a}. This kind of intermittency is negligible for saltation, since the transport rate of rolling grains is much smaller than that of saltating grains. However, it is important for bedload, where it is known to occur also below $\Theta_t$ \cite{Pahtzetal20a}. However, since the average flow cannot sustain the average motion of grains, such rolling events end very quickly below $\Theta_t$. Second, turbulent fluctuation events that temporarily push $\Theta^{\rm fluc}$ above $\Theta^{\rm cont}$ cause a different kind of intermittency, which is usually associated with saltation \cite{Comolaetal19b}, though the exact mechanism of this intermittency depends on the physical processes behind $\Theta^{\rm cont}$ \cite{Pahtzetal20a}. In the context of our proposed definition of $\Theta^{\rm cont}$, such events will cause grains entrained by grain-bed impacts to approach a periodic saltation trajectory, thus generating a saltation chain reaction that rapidly increases the transport rate $Q$. Once the turbulent fluctuation event is over, provided that $\Theta>\Theta_t$, saltation may maintain a large value of $Q$ for a relatively long time \cite{Pahtzetal20a}, though not indefinitely as trajectory fluctuations will cause grains to eventually settle (section~\ref{ThresholdInterpretation}). Nonetheless, this leads to a substantial hysteresis of $Q$ for $\Theta_t<\Theta<\Theta^{\rm cont}$ \cite{Carneiroetal15}.

Lastly, we emphasize that, although turbulence plays a crucial role for the complex intermittent behavior of NST for $\Theta_t<\Theta<\Theta^{\rm cont}$, both $\Theta_t$ and $\Theta^{\rm cont}$ are statistical quantities referring to the grain motion averaged over long times. That is, the influence of turbulence on $\Theta_t$ and $\Theta^{\rm cont}$ is probably relatively weak.

\subsection{Importance of Cohesion} \label{Cohesion}
Despite not explicitly accounting for cohesion, the coupled model captures measurements of the transport threshold $\Theta_t$ for small ($d\approx75~\mu\mathrm{m}$), that is, cohesive windblown sand grains by \citeA{Bagnold37} and \citeA{Chepil45} and measurements of $\Theta_t$ and the dimensionless transport rate $Q_\ast$ for potentially very cohesive windblown snow by \citeA{Sugiuraetal98}, as shown in Figures~\ref{Comparison}(c), \ref{Comparison}(d), and \ref{Comparison}(f). In particular, the model suggests that the increase of $\Theta_t$ with decreasing grain size $d$ for sufficiently small $d$, which was previously attributed to cohesion \cite<e.g.,>[]{Berzietal17,ShaoLu00}, is solely due to NST entering the viscous saltation regime. In this regime, $\Theta_t\propto(s^{1/2}Ga)^{-2}\sim d^{-3}$, which is a stronger decrease than the one ($\Theta_t\sim d^{-2}$) predicted by standard cohesion-based models \cite{Berzietal17,ShaoLu00}.

The agreement of the model with cohesive data supports the controversial suggestion by \citeA{Comolaetal19a} that, for equilibrium aeolian NST, $\Theta_t$ and $Q_\ast$ are nearly unaffected by the strength of cohesive bonds between bed grains. Further support comes from the model conceptualization. In fact, for the saltation regime, to which aeolian NST belongs, the only manner in which bed grains affect the model conceptualization is via the rebound laws, since both $Q_\ast$ \cite{PahtzDuran20} and $\Theta_t$ (section~\ref{Model}) have been conceptually introduced as bed sediment entrainment-independent physical quantities. However, the rebound laws are insensitive to the strength of cohesive bonds between bed grains for the simulation data by \citeA{Comolaetal19a} (Figures~\ref{Rebound}(d) and \ref{Rebound}(e)). Note that the DEM-based numerical simulations of equilibrium aeolian NST by \citeA{Comolaetal19a} were limited to Earth's atmospheric conditions, for which saltating grains perform relatively large hops on average \cite<$10\dots100d$, see>[Table~A.4]{Hoetal14}. However, our model conceptualization suggests that the insensitivity of equilibrium NST to cohesion is not limited to Earth's atmospheric conditions but applies to equilibrium NST in general as long as transported grains are unable to form cohesive bonds with bed grains during their contacts with the bed. This requirement is probably satisfied in the saltation regime (dashed lines in Figure~\ref{ModelApproximations}(b)) as the duration of grain-bed rebounds is very short. However, it may be violated in the bedload regime (solid lines in Figure~\ref{ModelApproximations}(b)) as grains tend to move in enduring contact with the bed. We therefore propose that the effects of cohesion tend to become negligible once the model predicts $\Theta^{\rm roll}_t=0$, which is the criterion with which we identify the saltation regime (Figure~\ref{ModelApproximations}(a)). However, we recognize that this proposition is controversial.

\section{Conclusions} \label{Conclusions}
In this study, we have proposed and evaluated a model of the two arguably most important statistical properties of equilibrium NST: the transport threshold Shields number $\Theta_t$ and dimensionless transport rate $Q_\ast$. The model captures, within a factor of about $2$, experimental and numerical data of $\Theta_t$ across the entire range of environmental conditions and experimental and numerical data of $Q_\ast$ across weak and intense conditions in which a small critical value of the Stokes-like number $s^{1/2}Ga$ (equation~(\ref{ValidityCriterion})) is exceeded (Figure~\ref{Comparison}). Such conditions include \textit{subaqueous bedload}, \textit{windblown sand}, and \textit{windblown snow}. The conditions that are not captured by the transport rate model correspond solely to \textit{viscous bedload} (Figure~\ref{ModelApproximations}(b)). Note that the agreement between model and experimental data includes the first controlled measurements of $Q_\ast$ for intense windblown sand by \citeA{Ralaiarisoaetal20}, which are not captured by standard expressions for $Q_\ast$ from the literature.

The agreement between the model and experimental data is not the result of parameter adjustment. In fact, all model parameters have been obtained from independent data sets: $c_M$ and $Z_\Delta$ from DEM-based numerical simulations of NST \cite{PahtzDuran18a,PahtzDuran20}, $A$, $B$, and $C$ from grain-bed rebound experiments \cite{Beladjineetal07}, and $\Theta^o_Y$ and $\psi_s$ from experiments and numerical simulations of bed yielding and grain entrainment driven by laminar flows \cite{Agudoetal17,Charruetal04,Houssaisetal15,Loiseleuxetal05,Ouriemietal07}.

NST is a highly fluctuating physical process \cite{Ancey20b,Duranetal11} because of turbulence, surface inhomogeneities, and variations of grain size and shape and packing geometry. Furthermore, the energy of transported grains varies strongly due to variations of their flow exposure duration since their entrainment from the bed \cite{Duranetal11}. However, such internal variability is completely neglected in the model, since it represents equilibrium NST by grains moving in identical periodic saltation trajectories. The high predictive capability of the model therefore suggests that crucial statistical properties of NST are relatively insensitive to its internal variability.

Although the model represents threshold conditions by a continuous grain motion, we have shown that $\Theta_t$ must be strictly smaller than the continuous transport threshold $\Theta^{\rm cont}$ for realistic natural settings (section~\ref{ThresholdInterpretation}). In particular, a semiempirical extension of the model for aeolian NST predicts values of $\Theta^{\rm cont}$ that are consistent with measurements (Figure~\ref{Stability}(e)). For $\Theta<\Theta^{\rm cont}$, NST can exhibit complex intermittency characteristics.

The model straightforwardly provides a criterion, which we validated with numerical data from the literature (Figure~\ref{ModelApproximations}(a)), that distinguishes \textit{bedload}, defined as NST in which a significant portion of transported grains is moving in enduring contacts with the bed surface (e.g., via rolling and sliding), from \textit{saltation}, defined as NST in which this portion is insignificant (section~\ref{TransportRegimes}). Based on the conceptualization of the model, we have proposed that, in the saltation regime, equilibrium NST is insensitive to the cohesiveness of bed grains (section~\ref{Cohesion}). Consistently, the cohesionless model captures measurements for aeolian saltation of cohesive sand and snow grains. In particular, the increase of $\Theta_t$ with decreasing grain size $d$ for windblown sand with $d\lesssim100~\mu\mathrm{m}$, previously attributed to cohesion \cite<e.g.,>[]{Berzietal17,ShaoLu00}, is predicted to be solely caused by NST entering the \textit{viscous saltation} regime (Figure~\ref{ModelApproximations}(b)), corresponding to saltation within the viscous sublayer of the turbulent boundary layer. In this regime, the model predicts $\Theta_t\propto(s^{1/2}Ga)^{-2}\sim d^{-3}$, which is a stronger decrease than the one ($\Theta_t\sim d^{-2}$) predicted by standard cohesion-based models \cite{Berzietal17,ShaoLu00}. Hence, the model supports the controversial notion that the strength of cohesive bonds between bed grains does neither significantly affect $\Theta_t$ nor $Q_\ast$, which \citeA{Comolaetal19a} suggested based on their DEM-based numerical simulations of equilibrium aeolian NST for Earth's atmospheric conditions.

Classically, the transport threshold has been corrected for a nonzero slope number $S$ via $\Theta_t=(1-S/\tan\alpha_r)\Theta_t|_{S=0}$, where $\alpha_r$ is the angle of repose \cite{IversenRasmussen94,Maurinetal18}. However, the model predicts that the predominant slope correction factor for turbulent NST is actually $1-S/\mu^o_b$ (equation~(\ref{STurbulentNST})). Although $\mu^o_b\simeq0.646$ is very close to typical values of $\tan\alpha_r$, its physical meaning in the model is fundamentally different. It is a purely kinematic bed friction coefficient associated with the laws that describe a grain-bed rebound. Furthermore, for viscous bedload, the model predicts a much milder bed slope dependency than the classical one (equation~(\ref{SViscousBedload})), in agreement with measurements (Figure~\ref{Slope}(d)). Note that the model predictions do not take into account that large bed slopes in nature (e.g., for mountain streams) are usually accompanied by very small flow depths of the order of $1d$, which cause bed mobility to decrease rather than increase with $S$~\cite{PrancevicLamb15,Prancevicetal14}.

Both the insensitivity of $\Theta_t$ to cohesion and the insensitivity of its slope dependency to $\alpha_r$ are features of the fact that $\Theta_t$ in the model is \textit{not} associated with the entrainment of bed surface grains, neither by the flow nor grain-bed impacts. Instead, $\Theta_t$ in the model can be characterized as a \textit{rebound threshold}, as defined and extensively discussed in a recent review on the matter \cite{Pahtzetal20a}.

In the future, the model may be used to reliably predict equilibrium NST in extraterrestrial environments, such as on Venus, Titan, Mars, and Pluto. However, while the model can probably be applied to Venus and Titan conditions, since they are well within the range of environmental conditions for which we evaluated the model, the application of the model to conditions with very large density ratios $s\gtrsim10^4$ (e.g., Mars and Pluto) requires further model validation. For this reason, DEM-based numerical simulations of NST for conditions with large $s$ are planned in the future.

\appendix
\section{Derivation of Equation~(\ref{M})} \label{MomentumBalances}
\citeA{PahtzDuran20} derived their equation~(S20), which is the analog to equation~(\ref{M}), from the fluid and particle momentum balances for equilibrium conditions:
\begin{linenomath*}
\begin{subequations}
\begin{align}
 \frac{\mathrm{d}\tau_f}{\mathrm{d}z}&=f_x,\quad\text{with}\quad\tau_f(z\geq z_{\rm max})=\tau, \label{FluidMomentum} \\
 \frac{\mathrm{d}\tau_g}{\mathrm{d}z}&=-f_x-\rho g\sin\alpha, \label{Momentumx} \\
 \frac{\mathrm{d}p_g}{\mathrm{d}z}&\simeq-\rho\tilde g, \label{Momentumz}
\end{align}
\end{subequations}
\end{linenomath*}
where $\tau_f$ is the fluid shear stress disturbed by the grain motion, $\tau_g$ the particle shear stress, $p_g$ the vertical particle pressure, $f_x$ the streamwise fluid-particle interaction force per unit volume (consisting of fluid drag and buoyancy in this paper), $\rho$ the particle concentration, and $z_{\rm max}$ the top of the transport layer, which no grain exceeds (i.e., $f_x=\rho=0$ for $z\geq z_{\rm max}$). Here, we argue for an improvement of equation~(\ref{FluidMomentum}), which is the streamwise fluid momentum balance using the inner turbulent boundary layer approximation.

Let us consider the fluid momentum balance, without boundary layer approximation, for slope-driven equilibrium NST. It reads \cite{Maurinetal18} 
\begin{linenomath*}
\begin{equation}
 \frac{\mathrm{d}\tau_f}{\mathrm{d}z}=-(\rho_f-s^{-1}\rho)g\sin\alpha+f_x,\quad\text{with}\quad\tau_f(z=h_m)=0, \label{FluidMomentum2}
\end{equation}
\end{linenomath*}
where $h_m$ is the elevation of the free surface relative to the bed surface elevation $z=0$, defined in Figure~\ref{Sketch} (i.e., $h_m$ is the flow depth, including the sediment-fluid mixture above the bed surface). Integrating equation~(\ref{FluidMomentum2}) from $z=0$ to $z=h_m$ yields
\begin{linenomath*}
\begin{equation}
 \tau-\tau_f(0)\equiv\rho_fgh_m\sin\alpha-\tau_f(0)=\int_0^{z_{\rm max}}(f_x+s^{-1}\rho g\sin\alpha)\mathrm{d}z. \label{FluidMomentum3}
\end{equation}
\end{linenomath*}
Hence, for the inner turbulent boundary layer approximation to be consistent with such slope-driven conditions, equation~(\ref{FluidMomentum}) should be modified to
\begin{linenomath*}
\begin{equation}
 \frac{\mathrm{d}\tau_f}{\mathrm{d}z}=f_x+s^{-1}\rho g\sin\alpha,\quad\text{with}\quad\tau_f(z\geq z_{\rm max})=\tau. \label{FluidMomentum4}
\end{equation}
\end{linenomath*}
Integrating equations~(\ref{Momentumx}), (\ref{Momentumz}), and (\ref{FluidMomentum4}) from $z=0$ to $z=z_{\rm max}$, assuming $\tau-\tau_f(0)\simeq\tau-\tau_t$ \cite{PahtzDuran20} and using $M\equiv\int_0^{z_{\rm max}}\rho\mathrm{d}z$, then yields
\begin{linenomath*}
\begin{subequations}
\begin{align}
 \tau-\tau_t\simeq\tau-\tau_f(0)&=\int_0^{z_{\rm max}}f_x\mathrm{d}z+s^{-1}Mg\sin\alpha, \\
 \tau_g(0)&=\int_0^{z_{\rm max}}f_x\mathrm{d}z+Mg\sin\alpha, \\
 p_g(0)&\simeq M\tilde g.
\end{align}
\end{subequations}
\end{linenomath*}
Combining these equations, using $\mu_b\equiv\tau_g(0)/p_g(0)$, yields equation~(\ref{M}).

\section{The Mean Flow Velocity Assumption} \label{MeanFlowVelocityAssumption}
We assume that the mean motion of grains driven by a fluctuating turbulent flow is the same as the mean motion of grains driven by a mean turbulent flow (section~\ref{FlowVelocityProfile}). To be approximately satisfied, this assumption has two requirements. First, turbulent diffusion must be negligible, otherwise, turbulent ejection events exercise a substantial control on the mean motion of grains \cite{AksamitPomeroy18a,Lelouveteletal09}. This requirement is satisfied because we consider only fully nonsuspended sediment transport \cite<i.e., Rouse number $Ro\gtrsim2.8$,>[]{Naqshbandetal17}. Second, the ratio between the standard deviation $\sigma_\tau$ and mean $\tau$ of the fluctuating applied fluid shear stress $\tau^{\rm fluc}$ should be sufficiently small so that streamwise grain velocity fluctuations are dominated by the randomness caused by interactions with the bed surface rather than fluid shear stress fluctuations. Note that we are modeling only those grains that have approached a nontrivial steady state trajectory, that is, comparably energetic grains that have survived multiple interactions with the bed surface without being captured. The velocity distribution of such grains would be expected to be Gaussian if their velocity fluctuations were predominantly caused by grain-bed interactions \cite{Hoetal12}, while it would be expected to be skewed (e.g., log-normal or exponential) if their velocity fluctuations were predominantly caused by streamwise flow fluctuations \cite{ShimDuan19} because $\tau^{\rm fluc}$ is log-normally distributed \cite{ChengLaw03,Martinetal13}. Coupled DEM/large eddy simulations by \citeA{Liuetal19} of a grain saltating along a fixed quadratically arranged (i.e., idealized) bed driven by water showed a skewed streamwise velocity distribution and, consistently, a substantial difference in the average streamwise grain motion when compared with simulations in which turbulent fluctuations were turned off. In contrast, experiments indicate symmetric Gaussian-like distributions of the streamwise velocity of energetic grains in natural aeolian and fluvial NST along random close packed (i.e., nonidealized) erodible beds \cite{AnceyPascal20,Heymanetal16,Kangetal08a,ShimDuan19}. We take this as evidence that, for natural conditions, the second requirement is approximately satisfied, though we recognize the potential of making a substantial error when assuming that the fluctuating turbulent flow can be approximated by its mean for modeling the mean motion of energetic grains.

\section{From Bed Friction Law to Periodic Saltation With Rebound Boundary Conditions} \label{SaltationJustification}
This appendix presents justifications, partially based on the bed friction law in section~\ref{FrictionLaw}, for why one can represent the entire grain motion in NST, including grains that roll and/or slide along the bed, by a periodic saltation motion with rebound boundary conditions. First, we justify representing NST by a pure saltation motion (section~\ref{JustificationSaltation}). Second, we justify the use of rebound boundary conditions (section~\ref{JustificationRebound}).

\subsection{Justification for Representing NST by a Pure Saltation Motion} \label{JustificationSaltation}
Like previous studies \cite<e.g.,>[]{Bagnold73,Charruetal04}, we assume that one can represent the entire grain motion in NST, including grains that roll and/or slide along the bed, by a pure saltation motion. A heuristic justification for this assumption is that bed grains that initially roll after their entrainment from the bed surface very quickly begin to make very small hops due to the geometrical disorder of the bed \cite<e.g., see>[Movie~S3]{Heymanetal16}. A more physical justification of this assumption was provided by \citeA{PahtzDuran18b} based on the bed friction law in section~\ref{FrictionLaw}. To explain these authors' findings, we introduce their precise definition of the bed friction coefficient: $\mu_b\equiv\tau_g(0)/p_g(0)$, where $\tau_g$ is the particle shear stress and $p_g$ the vertical particle pressure. Both $\tau_g\equiv\tau^c_g-\rho\langle v^\prime_zv^\prime_x\rangle$ and $p_g\equiv p^c_g+\rho\langle v_z^{\prime2}\rangle$ are separated into a contribution from particle contacts (superscript $c$) and a kinetic contribution associated with the grain fluctuation motion between contacts ($\pm\rho\langle v^\prime_zv^\prime_i\rangle$). \citeA{PahtzDuran18b} found that, for a wide range of DEM-based numerical simulations of NST, the kinetic contributions dominate when most grains move in saltation (typical for aeolian NST), while the contact contributions dominate when a significant portion of grains roll and/or slide along the bed (typical for fluvial NST). Nonetheless, these authors physically derived that, for near-threshold conditions, $\mu_b$ can always be approximated as the ratio between only the kinetic contributions, $\mu_b\approx-\langle v^\prime_zv^\prime_x\rangle(0)/\langle v_z^{\prime2}\rangle(0)$, and validated this derivation with all their near-threshold simulation data. This insight supports modeling NST across all regimes as a contact-free grain motion (i.e., saltation) above a flat wall, where grain-bed contacts are encoded in the wall boundary conditions.

\subsection{Justification for Using Rebound Boundary Conditions} \label{JustificationRebound}
Using rebound boundary conditions for a pure saltation motion is natural, since we are modeling only those grains that have approached a nontrivial steady state trajectory, that is, comparably energetic grains that have survived multiple rebounds with the bed surface without being captured. However, in regard to grains that roll and/or slide, additional justification is needed. A heuristic justification, based on the notion that a rolling regime is equivalent to a regime in which grains perform very small hops (section~\ref{JustificationSaltation}), is that any grain hopping along the surface will in time experience the entire phase space of possible impact conditions regardless of the typical size of its hops. That is, the impact conditions averaged over sufficiently many impacts are the same for grains performing large hops (saltation) or very small hops (rolling). The only difference is that grains performing large (very small) hops experience the same (varying) statistical impact conditions every single impact. However, we argue that this difference does not matter because we are only interested in modeling the grain dynamics averaged over all fluctuations.

To provide further justification for using rebound boundary conditions, we approximate $\overline{u_x(z)}\approx u_x(\overline{z})$ \cite{PahtzDuran18a} and rewrite equation~(\ref{Settling}) as
\begin{linenomath*}
\begin{equation}
  u_x(\overline{z})-\overline{v_x}=(\mu_b-S)v_s. \label{Settling2}
\end{equation}
\end{linenomath*}
\citeA{PahtzDuran18a} used equation~(\ref{Settling2}) in the limit of threshold conditions to predict the transport threshold $\Theta_t$. To do this, these authors derived $\overline{z}_t=z_{ct}+\overline{v_z^2}_t/\tilde g$, where $z_{ct}$ is the transport layer thickness due to particle contacts (unimportant for our discussion), and combined equation~(\ref{Settling2}) with equation~(\ref{uxcomplex}) and three semiempirical closures:
\begin{subequations}
\begin{linenomath*}
\begin{align}
 \mu_{bt}&=\mathrm{const}, \label{mubt} \\
 (\overline{v_z^2}_t)^{1/2}&\propto\overline{v_x}_t, \label{vzvx} \\
 \frac{\overline{v_x}_t}{u_{\ast t}}&=\frac{2}{\kappa}\sqrt{1-\exp\left[-\frac{1}{4}\kappa^2\gamma^2\left(\frac{\overline{u_x}_t}{u_{\ast t}}\right)^2\right]} \label{vxux},
\end{align}
\end{linenomath*}
\end{subequations}
where $\gamma$ is a constant. Using DEM-based simulations of NST for $S=0$, they found that these closures are roughly valid for near-threshold conditions with seemingly arbitrary $s$ and $Ga$ (i.e., including NST regimes with a significant or predominant rolling motion of grains), with the exception of equation~(\ref{vzvx}) for \textit{viscous saltation} (i.e., aeolian saltation within the viscous sublayer, a precise definition is provided in section~\ref{TransportRegimes}), and obtained the values of the parameters that appear in these closures from fitting to their simulation data.

Here, rather than from data fitting, we recover these closures, including the deviation for viscous saltation, from a periodic saltation model with rebound boundary conditions, supporting the use of such a model even for NST regimes with a significant or predominant rolling motion of grains. In fact, it is shown in section~\ref{ParticleBedRebounds} that equations~(\ref{mubt}) and (\ref{vzvx}) correspond to the rebound boundary conditions provided that vertical drag is negligible (i.e., $|a^d_z|\ll\tilde g$), which is significantly violated only for viscous saltation (causing slight deviations from equation~(\ref{mubt}) and substantial deviations from equation~(\ref{vzvx}), see section~\ref{TransportRegimes}). Furthermore, \citeA{PahtzDuran18a} showed that the two extreme regimes in equation~(\ref{vxux}), $\overline{v_x}_t\simeq\gamma \overline{u_x}_t$ for sufficiently small $\overline{u_x}_t/u_{\ast t}$ and $\overline{v_x}_t\simeq2\kappa^{-1}u_{\ast t}$ for sufficiently large $\overline{u_x}_t/u_{\ast t}$, follow from a minimization of $\Theta$. That is, the minimization of $\Theta$ in the present model (equation~(\ref{Thetat})) is the analog of equation~(\ref{vxux}).

\section{Fluid-Particle Interactions} \label{AppendixFluidParticleInteractions}
\subsection{Drag and lift}
\citeA{Lietal19} used highly resolved DNS simulations to measure the time series of the bed-tangential (drag) and bed-normal (lift) forces acting on a stationary spherical grain resting in a bed surface pocket or $d/3$ or $d$ above it. That is, these authors measured the total lift force comprised of turbulent lift and shear lift \cite{Deyetal20}. They found that this total lift force is negligible relative to the drag force (compare their Figures~9 and 10), except for a grain resting in the pocket. But even in this exceptional case (in most conditions that we model, the grain is most of the time a significant distance above the bed), the average lift force is still only about $1/3$ of the drag force. Further contributions to the lift force that would arise if the grain was in motion, such as Magnus lift and centrifugal lift \cite{Deyetal20}, are generally substantially smaller than that of the total shear and turbulent lift \cite{Zengetal09}. Owing to the fact that the average lift force rapidly decreases with the distance from the bed \cite{Chepil61} and even becomes negative \cite{Lietal19,Moragaetal99} in a poorly understood fashion, it would be a very difficult task to analytically account for it in a reliable manner even if we chose to not neglect it. Furthermore, even the analytical description of the drag force is associated with substantial uncertainties. For example, standard empirical expressions for the drag force substantially underestimate the actual drag force acting on a spherical grain resting in a bed surface pocket \cite{Lietal19}, while the form drag coefficient ($C^\infty_d$ in equation~(\ref{ad})) varies by a factor of about $4$ for typical grain shapes in nature \cite{Camenen07,Raffaeleetal20}. Given this large uncertainty, it makes little sense to consider additional forces of typically much lower magnitude than the drag force using further and even more uncertain empirical expressions (section~\ref{FluidParticleInteractions}).

\subsection{Basset force}
The Basset force depends on a grain's motion history and may become important when the shear Reynolds number $Re_d$ is sufficiently small, like for sand grains in water \cite{Bombardellietal08}. However, its magnitude relative to other fluid-particle interaction forces has been a matter of controversial debate \cite{Lukerchenko10}, since the Basset force contribution during a grain-bed interaction tends to compensate the one during the grain's contact-free motion \cite{Lukerchenko12}, especially for the steady grain motion considered in the model. For this reason, and since the Basset force is very difficult to be analytically incorporated, it is here neglected.

\subsection{Added mass force}
\citeA{PahtzDuran18a} have compared simulations with and without an added mass force term. The results were almost identical, even in the limit $s\rightarrow1$ where one would have expected otherwise. The reason is the fact that the added mass force is proportional to the sum of the total noncontact force (as it would be in the absence of the added mass effect) and the total contact force, which tend to compensate each other on average when $s$ is close to unity. Hence, if we included the added mass force in the transport threshold model, which does not account for contact forces for reasons explained in section~\ref{MathematicalModel}, it would only affect the fluid forces and thus create a spurious effect that is not real.

\section{Analytical Solution of Equations~(\ref{LawWall})-(\ref{az}) and its Approximation} \label{AppendixAnalyticalSolution}
From solving the differential equations~(\ref{LawWall})-(\ref{az}), with the initial condition $\mathbf{\hat v}(0)=\mathbf{\hat v_\uparrow}$, for $\mathbf{\hat v}(\hat t)$, the following expressions are obtained (which can be verified through insertion):
\begin{linenomath*}
\begin{subequations}
\begin{align}
 \hat z(\hat t)&=(1+\hat v_{\uparrow z})\left(1-e^{-\hat t}\right)-\hat t, \label{zt} \\
 \hat v_z(\hat t)&=(1+\hat v_{\uparrow z})e^{-\hat t}-1, \label{vzt} \\
 \hat x(\hat t)&=\hat v_{\uparrow x}\left(1-e^{-\hat t}\right)+S\left(\hat t+e^{-\hat t}-1\right)+\int_0^{\hat t}\int_0^{\hat t^{\prime\prime}}\hat u_x[\hat z(\hat t^\prime)]e^{\hat t^\prime-\hat t^{\prime\prime}}\mathrm{d}\hat t^\prime\mathrm{d}\hat t^{\prime\prime}, \label{xt} \\
\begin{split}
 \hat x(\hat t)&\approx\hat v_{\uparrow x}\left(1-e^{-\hat t}\right)+S\left(\hat t+e^{-\hat t}-1\right) \\
 &+\hat u_x\left\{\frac{2[3+\hat t+(2+\hat t)\hat v_{\uparrow z}]-e^{\hat t}[6+\hat t^2+4\hat v_{\uparrow z}-2\hat t(2+\hat v_{\uparrow z})]}{2+2e^{\hat t}(\hat t-1)}\right\}\left(\hat t+e^{-\hat t}-1\right), \label{xtapp}
\end{split}\\
 \hat v_x(\hat t)&=\hat v_{\uparrow x}e^{-\hat t}+S\left(1-e^{-\hat t}\right)+\int_0^{\hat t}\hat u_x[\hat z(\hat t^\prime)]e^{\hat t^\prime-\hat t}\mathrm{d}\hat t^\prime, \label{vxt} \\
 \hat v_x(\hat t)&\approx \hat v_{\uparrow x}e^{-\hat t}+S\left(1-e^{-\hat t}\right)+\hat u_x\left[2+\hat v_{\uparrow z}-\frac{\hat t\left(1+\hat v_{\uparrow z}+e^{\hat t}\right)}{e^{\hat t}-1}\right]\left(1-e^{-\hat t}\right). \label{vxtapp}
\end{align}
\end{subequations}
\end{linenomath*}
To eliminate the integrals, we have used the following approximation in equations~(\ref{xtapp}) and (\ref{vxtapp}):
\begin{linenomath*}
\begin{equation}
 \int_0^{\hat t}\hat u_x[f_1(\hat t^\prime)]f_2(\hat t^\prime)\mathrm{d}\hat t^\prime\approx\hat u_x\left[\frac{\int_0^{\hat t}f_1(\hat t^\prime)f_2(\hat t^\prime)\mathrm{d}\hat t^\prime}{\int_0^{\hat t}f_2(\hat t^\prime)\mathrm{d}\hat t^\prime}\right]\int_0^{\hat t}f_2(\hat t^\prime)\mathrm{d}\hat t^\prime, \label{IntegralApp}
\end{equation}
\end{linenomath*}
where the functions $f_1(\hat t)$ and $f_2(\hat t)$ stand representative for the functions within the inner integral in equation~(\ref{xt}), within the outer integral in equation~(\ref{xt}), or within the integral in equation~(\ref{vxt}). This approximation is exact for a linear velocity profile $\hat u_x(\hat z)$ (i.e., within the viscous sublayer) and very accurate within the log-layer, since $f_2(\hat t)$ changes much more rapidly with $\hat t$ than $\hat u_x[f_1(\hat t)]$ in the log-layer (regardless of the integral that is considered).

From equations~(\ref{zt}) and (\ref{vzt}), we obtain the nondimensionalized hop height $\hat H\equiv\hat z(\hat v_z=0)$ as
\begin{linenomath*}
\begin{equation}
 \hat H=\hat v_{\uparrow z}-\ln(1+\hat v_{\uparrow z}). \label{H}
\end{equation}
\end{linenomath*}
Furthermore, we obtain the nondimensionalized hop time $\hat T$ through setting $\hat z(\hat T>0)=0$ in equation~(\ref{zt}), which yields an implicit expression for $\hat T$:
\begin{linenomath*}
\begin{equation}
 \hat T=(1+\hat v_{\uparrow z})\left(1-e^{-\hat T}\right). \label{T}
\end{equation}
\end{linenomath*}
When comparing equation~(\ref{T}) with equation~(\ref{vzt}) evaluated at $\hat t=\hat T$, we obtain
\begin{linenomath*}
\begin{equation}
 \hat v_{\downarrow z}=\hat v_{\uparrow z}-\hat T. \label{viz1}
\end{equation}
\end{linenomath*}
Moreover, equation~(\ref{T}) can be rearranged to
\begin{linenomath*}
\begin{equation}
 \left[\hat T-(1+\hat v_{\uparrow z})\right]e^{\hat T-(1+\hat v_{\uparrow z})}=-(1+\hat v_{\uparrow z})e^{-(1+\hat v_{\uparrow z})}. \label{Thelp}
\end{equation}
\end{linenomath*}
Hence, using the definition of the principal branch of the Lambert-$W$ function (i.e., $Y=W(X)$ solves $X=Ye^Y$ for $Y\geq-1$) and $\hat T-(1+\hat v_{\uparrow z})=-\hat v_{\downarrow z}-1\geq-1$ (equation~(\ref{viz1})), we can solve equation~(\ref{Thelp}) for $\hat T$:
\begin{linenomath*}
\begin{equation}
 \hat T=1+\hat v_{\uparrow z}+W\left[-(1+\hat v_{\uparrow z})e^{-(1+\hat v_{\uparrow z})}\right]. \label{T2}
\end{equation}
\end{linenomath*}
Put together, equations~(\ref{viz1}) and (\ref{T2}) are equivalent to equation~(\ref{viz}).

Evaluating equations~(\ref{vxt}) and (\ref{vxtapp}) at $\hat t=\hat T$ yields
\begin{linenomath*}
\begin{subequations}
\begin{align}
 \hat v_{\downarrow x}-\hat v_{\uparrow x}&=\left\{-\hat v_{\uparrow x}+S+\int_0^{\hat T}\hat u_x[\hat z(\hat t)]\frac{e^{\hat t-\hat T}}{1-e^{-\hat T}}\mathrm{d}\hat t\right\}\left(1-e^{-\hat T}\right), \label{deltavx} \\
 \hat v_{\downarrow x}-\hat v_{\uparrow x}&\approx\left[-\hat v_{\uparrow x}+S+\hat u_x(\hat z_\ast)\right]\left(1-e^{-\hat T}\right),\text{ with}\quad\hat z_\ast\equiv-\hat v_{\downarrow z}(1+\hat v_{\uparrow z})-\hat v_{\uparrow z}, \label{deltavxapp}
\end{align}
\end{subequations}
\end{linenomath*}
where we used equations~(\ref{T}) and (\ref{viz1}) to obtain
\begin{linenomath*}
\begin{equation}
 2+\hat v_{\uparrow z}-\frac{\hat T\left(1+\hat v_{\uparrow z}+e^{\hat T}\right)}{e^{\hat T}-1}=-\hat v_{\downarrow z}(1+\hat v_{\uparrow z})-\hat v_{\uparrow z}.
\end{equation}
\end{linenomath*}
Furthermore, using equations~(\ref{mub}), (\ref{T}), and (\ref{viz1}) in equations~(\ref{deltavx}) and (\ref{deltavxapp}) yields after some rearrangement:
\begin{linenomath*}
\begin{subequations}
\begin{align}
 \int_0^{\hat T}\hat u_x[\hat z(\hat t)]\frac{e^{\hat t-\hat T}}{1-e^{-\hat T}}\mathrm{d}\hat t&=\mu_b(1+\hat v_{\uparrow z})+\hat v_{\uparrow x}-S, \label{uxmodel} \\
 \hat u_x(\hat z_\ast)&\approx\mu_b(1+\hat v_{\uparrow z})+\hat v_{\uparrow x}-S. \label{uxmodelapp}
\end{align}
\end{subequations}
\end{linenomath*}
It can be easily verified that equation~(\ref{uxmodelapp}) is equivalent to equation~(\ref{Thetacompapp}) after inserting equation~(\ref{LawWall}).

The average grain velocity-based quantities $\overline{\hat v_x}$ and $\overline{\hat v_z^2}$ are obtained from the definition of the time average ($\overline\cdot\equiv\frac{1}{\hat T}\int_0^{\hat T}\cdot\mathrm{d}\hat t$):
\begin{linenomath*}
\begin{subequations}
\begin{align}
 \overline{\hat v_x}&\approx\hat u_x(\overline{\hat z})-\mu_b+S, \label{meanvx} \\
 \overline{\hat z}=\overline{\hat v_z^2}&=\frac{1}{2}(\hat v_{\uparrow z}+\hat v_{\downarrow z}). \label{meanz}
\end{align}
\end{subequations}
\end{linenomath*}
For equation~(\ref{meanvx}), we used equation~(\ref{Settling}), $\overline{\hat v_x}\equiv\overline{v_x}/v_s$, and the approximation $\overline{\hat u_x}\approx\hat u_x(\overline{\hat z})$ (consistent with equation~(\ref{IntegralApp})), while equation~(\ref{meanz}) is derived from equations~(\ref{zt}), (\ref{vzt}), and (\ref{viz1}). Note that an alternative approximation for $\overline{\hat v_x}$, which yields almost the same values, is given by $\overline{\hat v_x}=\hat x(\hat T)/\hat T$ when using the approximation in equation~(\ref{xtapp}) to calculate $\hat x(\hat T)$.

\section{Transport Threshold Model in the Limit of Negligible Vertical Drag} \label{AppendixNoVerticalDrag}
The limit of negligible vertical drag (i.e., $|a^d_z|\ll\tilde g$) is equivalent to the limit in which grain velocities are much smaller than the terminal settling velocity $v_s$. Hence, we obtain this limit from Taylor-expanding the transport threshold model in leading order of $\hat T=T\tilde g/v_s$. For equations~(\ref{viz}) and (\ref{viz1}), this implies
\begin{linenomath*}
\begin{align}
 \hat v_{\downarrow z}&\simeq-\hat v_{\uparrow z}, \\
 \hat T&\simeq2\hat v_{\uparrow z},
\end{align}
\end{linenomath*}
which means that a grain's kinetic energy is conserved with respect to its vertical motion (i.e., $e_z\equiv-v_{\uparrow z}/v_{\downarrow z}\simeq1$), exactly as one would expect if gravity and buoyancy dominate. Using $e_z\simeq1$, it follows from equations~(\ref{e}) and (\ref{ez}) that all components of the dimensionless rebound and impact velocities are proportional to $\hat T$ and each other in this limit: $\hat v_{\uparrow z}\propto\hat v_{\downarrow z}\propto\hat v_{\uparrow x}\propto\hat v_{\downarrow x}\propto\hat T$. Hence, the Taylor expansions of equations~(\ref{uxmodel}) and (\ref{uxmodelapp}) in leading order of $\hat T$ read
\begin{linenomath*}
\begin{subequations}
\begin{align}
 \frac{1}{\hat T}\int_0^{\hat T}\hat u_x[\hat z(\hat t)]\mathrm{d}\hat t\equiv\overline{\hat u_x}&\simeq\mu^o_b(1+\hat v_{\uparrow z})+\hat v_{\uparrow x}-S,\text{ with }\hat z(\hat t)\simeq\hat v_{\uparrow z}\hat t-\frac{1}{2}\hat t^2, \label{Thetanodrag} \\
 \hat u_x(\overline{\hat z})&\approx\mu^o_b(1+\hat v_{\uparrow z})+\hat v_{\uparrow x}-S,\text{ with }\overline{\hat z}=\overline{\hat v_z^2}\simeq\frac{1}{3}\hat v_{\uparrow z}^2, \label{uxmodelappnodrag}
\end{align}
\end{subequations}
\end{linenomath*}
where $\mu^o_b$ can be calculated as (using equations~(\ref{e}) and (\ref{ez}))
\begin{linenomath*}
\begin{equation}
\begin{split}
 \mu^o_b&\equiv\lim\limits_{e_z\rightarrow1}\mu_b=\frac{1}{2(A+C)^2}\left\{\sqrt{(1+B+C)^4-(A+C)^4}\right. \\
 &\left.-\sqrt{[A(1+B+C)^2-B(A+C)^2]^2-(A+C)^4}\right\}\simeq0.646.
\end{split}
\end{equation}
\end{linenomath*}
Note that, using equation~(\ref{meanvx}), equation~(\ref{Thetanodrag}) implies
\begin{linenomath*}
\begin{equation}
 \overline{\hat v_x}\simeq\mu^o_b\hat v_{\uparrow z}+\hat v_{\uparrow x}=\frac{1}{2}(\hat v_{\uparrow x}+\hat v_{\downarrow x}). \label{vxapp}
\end{equation}
\end{linenomath*}

Interestingly, in the limit of negligible vertical drag, the dependency of the equations that describe the grain trajectory on the slope number $S$ can be directly linked to the case $S=0$. To show this, we define $f_S\equiv1-S/\mu^o_b$, $\mathbf{v_{\uparrow(\downarrow)}}|_{\ast S}\equiv f_S^{-1/2}\mathbf{v_{\uparrow(\downarrow)}}/\sqrt{s\tilde gd}$, $v^o_{s\ast}\equiv\lim_{e_z\rightarrow1}v_{s\ast}$, and the new dimensionless numbers $Ga_S\equiv f_S^{1/2}Ga$, $s_S\equiv f_Ss$, and $\Theta_S\equiv f_S^{-1}\Theta$. Then, using the definition of the hat ($\hat\cdot$) and equations~(\ref{Settling}) and (\ref{LawWall}), we can rewrite equation~(\ref{uxmodelappnodrag}) as
\begin{linenomath*}
\begin{equation} \label{SlopeNoDrag}
\begin{split}
 &\sqrt{\Theta_S}f_u\left(Ga_S\sqrt{\Theta_S},\frac{1}{3}s_Sv_{\uparrow z}^2|_{\ast S}\right)\simeq\mu^o_bv_{\uparrow z}|_{\ast S}+v_{\uparrow x}|_{\ast S}+\mu^o_b[\sqrt{f_S}v^o_{s\ast}](Ga_S),\text{ with} \\
 &[\sqrt{f_S}v^o_{s\ast}](Ga_S)=\frac{1}{\mu^o_b}\left[\sqrt{\frac{1}{4}\sqrt[m]{\left(\frac{24}{C_d^\infty Ga_S}\right)^2}+\sqrt[m]{\frac{4\mu^o_b}{3C_d^\infty}}}-\frac{1}{2}\sqrt[m]{\frac{24}{C_d^\infty Ga_S}}\right]^m.
\end{split}
\end{equation}
\end{linenomath*}
Equation~(\ref{SlopeNoDrag}) does not contain explicit dependencies on $S$, implying that the only remaining threshold model equation that contains an explicit dependency on $S$ is the one for $\Theta^{\rm roll}$ (equation~(\ref{Eb})). Furthermore, for $S=0$ (i.e., $f_S=1$), the functional forms of equation~(\ref{SlopeNoDrag}) remains unchanged if the modified quantities with subscript $S$ are replaced by the unmodified ones without subscript $S$. Hence, for saltation conditions ($\Theta^{\rm roll}_t\equiv\Theta^{\rm roll}|_{\Theta=\Theta_t}=0$) in the limit of negligible vertical drag, like for turbulent saltation, $\Theta_t$ and $\overline{v_x}_{\ast t}$ scale as
\begin{linenomath*}
\begin{align}
 \Theta_t(Ga,s,S)&\simeq f_S\Theta_t(\sqrt{f_S}Ga,f_Ss,0), \\
 \overline{v_x}_{\ast t}(Ga,s,S)&\simeq\sqrt{f_S}\overline{v_x}_{\ast t}(\sqrt{f_S}Ga,f_Ss,0).
\end{align}
\end{linenomath*}
It turns out that these scaling laws also work reasonably well for turbulent bedload even though $\Theta^{\rm roll}_t>0$ (Figure~\ref{Slope}(b)). Note that, for the derivation of equations~(\ref{Qapp}) and (\ref{Mapp}), we further roughly approximated the right-hand side of these expressions using $\Theta_t(f_S^{1/2}Ga,f_Ss,0)\approx\Theta_t(Ga,s,0)$ and $\overline{v_x}_{\ast t}(f_S^{1/2}Ga,f_Ss,0)\approx\overline{v_x}_{\ast t}(Ga,s,0)$, respectively, since the indirect effects of $f_S$ on $\Theta_t$ and $\overline{v_x}_{\ast t}$ via rescaling $Ga$ and $s$ are relatively weak.

\section{Explicit Solution of Equation~(\ref{ThetatTurbulentSaltation})} \label{AppendixTurbulentSaltation}
Equation~(\ref{ThetatTurbulentSaltation}) can be rewritten in the form
\begin{linenomath*}
\begin{equation}
 \Theta_t\simeq a[\ln(b\Theta_t)-c]^{-2}, \label{ThetatTurb1}
\end{equation}
\end{linenomath*}
where $a\equiv[\kappa(\mu^o_b-S)v^o_{s\ast}]^2$, $b\equiv30\kappa^{-2}\beta s$, and $c=2$ for $z_{ot}=d/30$ and $a\equiv[2\kappa(\mu^o_b-S)v^o_{s\ast}/3]^2$, $b\equiv(9\kappa^{-2}\beta sGa)^{2/3}$, and $c=4/3$ for $z_{ot}=d/(9Re_{dt})=d/(9\Theta_t^{1/2}Ga)$. Note that, using equations~(\ref{e}) and (\ref{ez}), the constant $\beta$ appearing in $b$ can be calculated as
\begin{linenomath*}
\begin{equation}
\begin{split}
 \beta&\equiv\lim\limits_{e_z\rightarrow1}\frac{4v_{\uparrow z}^2}{3(\mu_bv_{\uparrow z}+v_{\uparrow x})^2}=\frac{16}{3}(A+C)^4\left\{\sqrt{(1+B+C)^4-(A+C)^4}\right. \\
 &\left.+\sqrt{[A(1+B+C)^2-B(A+C)^2]^2-(A+C)^4}\right\}^{-2}\simeq0.136.
\end{split}
\end{equation}
\end{linenomath*}
Now, further rearrangement of equation~(\ref{ThetatTurb1}) successively yields
\begin{linenomath*}
\begin{align}
 \frac{\ln(b\Theta_t)-c}{2}\exp\left(\frac{\ln(b\Theta_t)-c}{2}\right)&\simeq\frac{\sqrt{ab}}{2e^{c/2}}, \\
 \frac{\ln(b\Theta_t)-c}{2}&\simeq W\left(\frac{\sqrt{ab}}{2e^{c/2}}\right), \\
 \Theta_t\simeq\frac{1}{b}\exp\left[2W\left(\frac{\sqrt{ab}}{2e^{c/2}}\right)+c\right]&=\frac{a}{4}W\left(\frac{\sqrt{ab}}{2e^{c/2}}\right)^{-2}, \label{ThetatTurb2}
\end{align}
\end{linenomath*}
where we used the definition of the principal branch of the Lambert-$W$ function and the identity $\exp[2W(X)]=[X/W(X)]^2$. The explicit solution for $\Theta_t$ in equation~(\ref{ThetatTurb2}) corresponds to $\Theta^{\rm rough}_t$ in equation~(\ref{ThetatTurbulentSaltation2}) for $z_{ot}=d/30$ and $\Theta^{\rm smooth}_t$ in equation~(\ref{ThetatTurbulentSaltation2}) for $z_{ot}=d/(9Re_{dt})$.

\acknowledgments
All data shown in the figures of this article can be found in the following references: \citeA{Comolaetal19a}, \citeA{PahtzDuran18a}, \citeA{PahtzDuran20}, \citeA{MeyerPeterMuller48}, \citeA{Karahan75}, \citeA{FernandezLuqueVanBeek76}, \citeA{YalinKarahan79}, \citeA{SmartJaeggi83}, \citeA{Loiseleuxetal05}, \citeA{Ouriemietal07}, \citeA{CapartFraccarollo11}, \citeA{Bagnold37}, \citeA{Chepil45}, \citeA{Creysselsetal09}, \citeA{Hoetal11}, \citeA{Ho12}, \citeA{Carneiroetal15}, \citeA{MartinKok17,MartinKok18}, \citeA{Zhuetal19}, \citeA{Ralaiarisoaetal20}, \citeA{Sugiuraetal98}, \citeA{Cliftonetal06}, and \citeA{Andreottietal21}. We thank four anonymous reviewers for their critical comments that helped improving the paper. We acknowledge support from grants National Natural Science Foundation of China (Nos.~42076177, 11772300, and 11672267), Key Research and Development Project of Zhejiang Province (2021C03180), and Zhejiang Natural Science Foundation (LR19E090002).

\begin{notation}
 \notation{$\kappa=0.4$} Von K\'arm\'an constant
 \notation{$C^\infty_d=0.4$, $m^d=2$} Constants associated with fluid drag law
 \notation{$c_M=1.7$} Model parameter associated with grain fluctuation energy dissipation
 \notation{$z_\Delta$} Distance between rebound elevation and virtual zero-level of flow velocity profile
 \notation{$Z_\Delta=0.7$} $\equiv z_\Delta/d$ (model parameter)
 \notation{$\Delta z_s$} Distance between bed crest level and virtual zero-level of flow velocity profile
 \notation{$A=0.87$, $B=0.72$, $C=0$} Model parameters associated with rebound laws
 \notation{$\Theta_Y$} Yield stress
 \notation{$\Theta^o_Y=0.13$} Yield stress for nonsloped bed ($S=0$)
 \notation{$\psi$} Pocket angle of a pocket of the bed surface
 \notation{$\psi_s=25^\circ$} Pocket angle of most stable bed surface pocket
 \notation{$\psi_\ast$} Pocket angle after kinetic energy of grain located in most stable bed surface pocket has been fully converted into potential energy during initial rolling motion
 \notation{$h_s$, $h_\ast$} Distances depicted in Figure~\ref{Escape}
 \notation{$\tau_g$} Particle shear stress
 \notation{$p_g$} Vertical particle pressure
 \notation{$\mu_b$} $\equiv\tau_g(0)/p_g(0)$ (bed surface value of friction coefficient)
 \notation{$\mu^o_b$} $\equiv\lim_{e_z\rightarrow1}\mu_b\simeq0.646$
 \notation{$\beta$} $\equiv\lim_{e_z\rightarrow1}(4v_{\uparrow z}^2/3)/(\mu_bv_{\uparrow z}+v_{\uparrow x})^2\simeq0.136$
 \notation{$\mathbf{u}$} Mean flow velocity
 \notation{$m$} Mass of single grain
 \notation{$\mathbf{v}$} Grain velocity
 \notation{$\mathbf{v^\prime}$} Grain fluctuation velocity
 \notation{$\mathbf{v_\uparrow}$} Lift-off or rebound velocity
 \notation{$\mathbf{v_\downarrow}$} Impact velocity
 \notation{$\mathbf{v^o_\uparrow}$} Initial lift-off or rebound velocity
 \notation{$\mathbf{v^e_\uparrow}$} Average velocity of entrained grain
 \notation{$\mathbf{E_\uparrow}$} $=\frac{1}{2}m\mathbf{v_\uparrow}^2$ (lift-off or rebound grain kinetic energy)
 \notation{$e$} $\equiv|\mathbf{v_\downarrow}|/|\mathbf{v_\uparrow}|$ (rebound restitution coefficient)
 \notation{$e_z$} $\equiv-v_{\uparrow z}/v_{\downarrow z}$ (vertical rebound restitution coefficient)
 \notation{$\theta_\uparrow$} Lift-off or rebound angle
 \notation{$\theta_\downarrow$} Impact angle
 \notation{$\mathbf{a}$} Grain acceleration
 \notation{$\mathbf{a^d}$} Grain acceleration due to fluid drag
 \notation{$\mathbf{a^g}$} Grain acceleration due to gravity
 \notation{$\mathbf{a^b}$} Grain acceleration due to buoyancy
 \notation{$f_x$} Streamwise fluid-particle interaction force per unit volume
 \notation{$z_{\rm max}$} Top of the transport layer, which no grain exceeds
 \notation{$h_m$} Flow depth, including the sediment-fluid mixture above the bed surface $z=0$
 \notation{$H$} Hop height in periodic saltation model
 \notation{$T$} Hop time in periodic saltation model
 \notation{$\rho$} Particle concentration
 \notation{$\rho^\uparrow$} Particle concentration of ascending grains
 \notation{$\rho^\downarrow$} Particle concentration of descending grains
 \notation{$\phi^\uparrow$} Vertical upward-flux of grains
 \notation{$G^\uparrow$} Quantity $G$ during upward-motion of periodic saltation trajectory
 \notation{$G^\downarrow$} Quantity $G$ during downward-motion of periodic saltation trajectory
 \notation{$\langle G\rangle$} Local grain mass-weighted ensemble average of quantity $G$
 \notation{$\overline{G}$} $\rho$-weighted height average of quantity $G$ (= time average over grain hop in periodic saltation model)
 \notation{$\hat{G}$} Quantity $G$ nondimensionalized using units of $v_s$ and $\tilde g$
 \notation{$G_t$} Quantity $G$ in the limit of threshold conditions
 \notation{$v^-_s$} Terminal grain settling velocity in quiescent flow
 \notation{$v_s$} Terminal grain settling velocity in equilibrium NST
 \notation{$v_{s\ast}$} Nondimensionalized terminal settling velocity of grains in equilibrium NST
 \notation{$v^o_{s\ast}$} $=\lim_{e_z\rightarrow1}v_{s\ast}$
 \notation{$d$} Grain diameter
 \notation{$\rho_p$} Particle density
 \notation{$\rho_f$} Fluid density
 \notation{$\nu_f$} Kinematic fluid viscosity
 \notation{$\tau$} Fluid shear stress applied onto bed surface
 \notation{$\tau^{\rm fluc}$} Fluctuating fluid shear stress applied onto bed surface
 \notation{$\sigma_\tau$} Standard deviation of fluctuating bed fluid shear stress
 \notation{$u_\ast$} Fluid shear velocity
 \notation{$\alpha$} Bed slope angle (slopes aligned with flow direction, for downward slopes, $\alpha>0$)
 \notation{$\alpha_r$} Angle of repose
 \notation{$f_S$} $\equiv1-S/\mu^o_b$ (slope correction factor)
 \notation{$f_\alpha$} $\equiv1-\tan\alpha/\mu^o_b$
 \notation{$g$} Gravity constant
 \notation{$\tilde g$} Vertical buoyancy-reduced value of $g$
 \notation{$f_u$, $f_{\tilde u}$} Functions associated with mean flow velocity profile
 \notation{$Ro$} $\equiv v^-_s/(\kappa u_\ast)$ (Rouse number, fully NST requires $Ro\gtrsim Ro_c=2.8$)
 \notation{$Re_d$} $\equiv u_\ast d/\nu_f$ (particle Reynolds number)
 \notation{$Re_z$} $\equiv Re_d(z/d+Z_\Delta)$ (wall units)
 \notation{$s$} $\equiv\rho_p/\rho_f$ (density ratio)
 \notation{$Ga$} $\equiv d\sqrt{s\tilde gd}/\nu_f$ (Galileo number)
 \notation{$S$} $\equiv-(a^g_x+a^b_x)/(a^g_z+a^b_z)$ (slope number)
 \notation{$\Theta$} $\equiv\tau/(\rho_p\tilde gd)$ (Shields number)
 \notation{$\Theta_t$} Shields number at transport threshold
 \notation{$\Theta^{\rm fluc}$} $\equiv\tau^{\rm fluc}/(\rho_p\tilde gd)$
 \notation{$\Theta^c$} Bed surface pocket angle-dependent critical Shields number required for rolling
 \notation{$\Theta^{\rm roll}$} Grain kinetic energy-dependent critical Shields number required for rolling
 \notation{$\Theta^{\rm cont}$} Shields number at continuous transport threshold
 \notation{$Q$} Sediment transport rate
 \notation{$Q_\ast$} $\equiv Q/(\rho_pd\sqrt{s\tilde gd})$
 \notation{$M$} Sediment transport load
 \notation{$M_\ast$} $\equiv M/(\rho_pd)$
 \notation{$\overline{v_x}$} Average streamwise sediment velocity
 \notation{$\overline{v_x}_\ast$} $\equiv\overline{v_x}/\sqrt{s\tilde gd}$
 \notation{$W$} Lambert-$W$ function
\end{notation}


\end{document}